\documentclass[aps,twocolumn,floatfix,showpacs,amsmath,amsfonts]{revtex4}

\usepackage{graphicx}
\usepackage{bm}
\usepackage{units}
\usepackage{color}
\usepackage{upgreek}

\begin{document}

\title{Magnetoexcitons in cuprous oxide}

\author{Frank Schweiner}
\author{J\"org Main}
\author{G\"unter Wunner}
\affiliation{Institut f\"ur Theoretische Physik 1, Universit\"at Stuttgart,
  70550 Stuttgart, Germany}
\author{Marcel Freitag}
\author{Julian Heck\"otter}
\author{Christoph Uihlein}
\author{Marc A{\ss}mann}
\author{Dietmar Fr\"ohlich}
\author{Manfred Bayer}
\affiliation{Experimentelle Physik 2, Technische Universit\"at Dortmund, 44221 Dortmund, Germany}
\date{\today}

\begin{abstract}
Two of the most striking experimental
findings when investigating exciton spectra in cuprous oxide using high-resolution
spectroscopy are the observability and the fine structure splitting of $F$ excitons
reported by J.~Thewes \emph{et~al}.
[Phys.~Rev.~Lett.~\textbf{115}, 027402 (2015)]. 
These findings show that it is indispensable to 
account for the complex valence band structure
and the cubic symmetry of the solid
in the theory of excitons.
This is all the more important for magnetoexcitons, where the external
magnetic field reduces the symmetry of the system even further.
We present the theory of excitons in $\mathrm{Cu_{2}O}$ in an 
external magnetic field and especially discuss the dependence of the spectra on the 
direction of the external magnetic field,
which cannot
be understood from a simple hydrogen-like model.
Using high-resolution
spectroscopy, we also present the corresponding experimental spectra for cuprous oxide
in Faraday configuration.
The theoretical results and experimental spectra are in excellent 
agreement as regards not only the energies but also the relative oscillator strengths.
Furthermore, this comparison allows for the
determination of the fourth Luttinger parameter $\kappa$ of this semiconductor. 
\end{abstract}


\maketitle

\section{Introduction}

Excitons are of great physical interest since they 
represent the
fundamental optical excitation in semiconductors.
Excitons in cuprous oxide 
$\left(\mathrm{Cu_{2}O}\right)$, in particular,
have attracted lots of attention in recent
years~\cite{GRE,QC,75,76,50,28,80,100,78,79,150,175,74,77}
due to an experiment,
in which the hydrogen-like absorption
spectrum of these quasi-particles could be 
observed up to a principal
quantum number of $n=25$~\cite{GRE}.
The discovery of these giant Rydberg excitons
may pave the way to a deeper
understanding of inter-particle interactions in the solid~\cite{GRE}
and to applications in quantum
information technology~\cite{77}.

In this context it is indispensable 
to completely understand the underlying
theory of excitons.
Excitons consist of a negatively 
charged electron in the conduction band 
and a positively charged hole in the valence band.
As the interaction between electron and hole
can be described by a screened Coulomb interaction,
excitons are often regarded as the solid-state analog
of a hydrogen atom~\cite{TOE_5,NM5_7,TOE,NM5_9}.
However, the hydrogen-like model of excitons 
is generally too simple to describe exciton spectra correctly.
It has recently been shown that even without external
fields this model is incapable of describing the
fine structure splitting observed experimentally
and that it is inevitable to account for
the complex valence band structure and the
cubic symmetry $O_{\mathrm{h}}$ of $\mathrm{Cu_{2}O}$
in a quantitative 
theory~\cite{28,100}.

This is all the more important in the presence of
an external magnetic field, which reduces the
symmetry of $\mathrm{Cu_{2}O}$ 
to a lower symmetry.
For this reason one expects an extremely complex splitting of
exciton lines in absorption spectra,
in which also anticrossings appear.
Hence, earlier theoretical 
treatments of these spectra using a
hydrogen-like model were unable to describe the 
vast number of lines observed in experiments 
(see Refs.~\cite{42,12,29a,29b} and further references therein).
On the other hand, due to the specific material parameters in 
$\mathrm{Cu_{2}O}$, the exciton radius $a_{\mathrm{exc}}$ 
is much larger than the Bohr radius $a_{0}$ known from atomic physics. 
This makes excitons attractive for 
investigations in external fields~\cite{50} since 
the region of ``high magnetic fields'' can be
reached within several Tesla, in contrast to the hydrogen atom, where
this region begins above several hundreds of Tesla~\cite{50,12}.

We present the theory
for the exciton absorption spectra of $\mathrm{Cu_{2}O}$
in an external magnetic field and solve the 
corresponding Schr\"odinger equation using a complete basis. 
This method also allows for the direct calculation of relative oscillator strengths.
We especially discuss the dependence of the spectra on the 
direction of the external magnetic field,
which is well described by the anisotropic band structure and
which cannot be understood from a simple hydrogen-like model.
Using high resolution spectroscopy and
natural crystals, we also present the complex
experimental absorption spectra for the $n\leq 7$ exciton states
in Faraday configuration
with a significantly better resolution 
than in previous work on this topic~\cite{42,12,29a,29b}.
The comparison of theory and experiment shows an excellent agreement.
It furthermore allows
for the determination of an yet not precisely determined material parameter of 
$\mathrm{Cu_{2}O}$, i.e., the fourth Luttinger parameter $\kappa$.

The paper is organized as follows:
In Sec.~\ref{sec:Hamiltonian} we present the Hamiltonian of
excitons in external fields in the case of degenerate valence bands and
the method of solving the Schr\"odinger equation in a complete basis.
We explain how to calculate relative oscillator strengths in Sec.~\ref{sec:Oscillator}.
In Sec.~\ref{sec:Experimental} the
experimental setup is described and the absorption spectra 
of excitons in a uniform magnetic field are discussed.
In Sec.~\ref{sub:field} we investigate the symmetry of the Hamiltonian and 
compare theoretical with experimental spectra for different orientations of the magnetic field
to determine the fourth Luttinger parameter of $\mathrm{Cu_{2}O}$.
Finally, we give a short summary and outlook in Sec.~\ref{sec:Summary}.

\section{Hamiltonian \label{sec:Hamiltonian}}

In this section we present the theory of
exciton spectra of $\mathrm{Cu_{2}O}$ in a uniform magnetic field.
The lowest conduction band in $\mathrm{Cu_{2}O}$
is almost parabolic in the vicinity of the $\Gamma$ point or
the center of the first Brillouin zone.
Therefore, the kinetic energy of
an electron in this $\Gamma_6^+$ 
conduction band is given by
\begin{equation}
H_{\mathrm{e}}\left(\boldsymbol{p}_{\mathrm{e}}\right)=\frac{\boldsymbol{p}_{\mathrm{e}}^{2}}{2m_{\mathrm{e}}}
\end{equation}
with the effective electron mass $m_{\mathrm{e}}$.
Since $\mathrm{Cu_{2}O}$ has cubic symmetry, we use the
irreducible representations $\Gamma_{i}^{\pm}$
of the cubic group $O_{\mathrm{h}}$ to assign the symmetry of the bands.
In contrast to the conduction band, 
the three uppermost valence bands in $\mathrm{Cu_{2}O}$
are nonparabolic but deformed due to interband interactions
and the presence of the non-spherical symmetry of the solid.
Hence, the kinetic energy of the hole is given 
by the more complex expression~\cite{80,100}
\begin{eqnarray}
H_{\mathrm{h}}\left(\boldsymbol{p}_{\mathrm{h}}\right) & = & H_{\mathrm{so}}+\left(1/2\hbar^{2}m_{0}\right)\left\{ \hbar^{2}\left(\gamma_{1}+4\gamma_{2}\right)\boldsymbol{p}_{\mathrm{h}}^{2}\right.\phantom{\frac{1}{1}}\nonumber \\
 & + & 2\left(\eta_{1}+2\eta_{2}\right)\boldsymbol{p}_{\mathrm{h}}^{2}\left(\boldsymbol{I}\cdot\boldsymbol{S}_{\mathrm{h}}\right)\phantom{\frac{1}{1}}\nonumber \\
 & - & 6\gamma_{2}\left(p_{\mathrm{h}1}^{2}\boldsymbol{I}_{1}^{2}+\mathrm{c.p.}\right)-12\eta_{2}\left(p_{\mathrm{h}1}^{2}\boldsymbol{I}_{1}\boldsymbol{S}_{\mathrm{h}1}+\mathrm{c.p.}\right)\phantom{\frac{1}{1}}\nonumber \\
 & - & 12\gamma_{3}\left(\left\{ p_{\mathrm{h}1},p_{\mathrm{h}2}\right\} \left\{ \boldsymbol{I}_{1},\boldsymbol{I}_{2}\right\} +\mathrm{c.p.}\right)\phantom{\frac{1}{1}}\nonumber \\
 & - & \left.12\eta_{3}\left(\left\{ p_{\mathrm{h}1},p_{\mathrm{h}2}\right\} \left(\boldsymbol{I}_{1}\boldsymbol{S}_{\mathrm{h}2}+\boldsymbol{I}_{2}\boldsymbol{S}_{\mathrm{h}1}\right)+\mathrm{c.p.}\right)\right\} \phantom{\frac{1}{1}}\label{eq:Hh}
\end{eqnarray}
with $\boldsymbol{p}=\left(p_1,\,p_2,\,p_3\right)$, $\left\{ a,b\right\} =\frac{1}{2}\left(ab+ba\right)$ and
c.p.~denoting cyclic permutation.
The parameters $\gamma_{i}$ and $m_{0}$ denote the first three Luttinger parameters
and the free electron mass, respectively. The parameters $\eta_{i}$ are much smaller than the Luttinger parameters~\cite{100}.
All of these coefficients describe the behavior and the 
anisotropic effective mass of the hole in the vicinity of the $\Gamma$ point.
The matrices $\boldsymbol{I}_j$ and $\boldsymbol{S}_{\mathrm{h}j}$ denote the three
spin matrices of the quasispin $I=1$ and the hole spin $S_{\mathrm{h}}=1/2$
while $\boldsymbol{I}$ and $\boldsymbol{S}_{\mathrm{h}}$ are vectors containing
these matrices, i.e.,
\begin{equation}
\boldsymbol{I}\cdot\boldsymbol{S}_{\mathrm{h}}=\sum_{j=1}^{3}\boldsymbol{I}_{j}\boldsymbol{S}_{\mathrm{h}j}
\end{equation}
The quasispin $I=1$
describes the threefold degenerate valence band and
is a convenient abstraction to denote the three 
orbital Bloch functions $xy$, $yz$, and $zx$~\cite{25}.
Due to the spin-orbit coupling
between the quasispin $I$ and the hole spin $S_{h}$~\cite{7}
\begin{equation}
H_{\mathrm{so}}=\frac{2}{3}\Delta\left(1+\frac{1}{\hbar^{2}}\boldsymbol{I}\cdot\boldsymbol{S}_{\mathrm{h}}\right),\label{eq:soc}
\end{equation}
the sixfold degenerate valence band (including the hole spin) splits
into a higher lying twofold-degenerate band of symmetry $\Gamma_{7}^{+}$
and a lower lying fourfold-degenerate band of symmetry $\Gamma_{8}^{+}$
by an amount of $\Delta$.

The Hamiltonian of the exciton is then
given by~\cite{17_17,7}
\begin{equation}
H=E_{\mathrm{g}}+V\left(\boldsymbol{r}_{e}-\boldsymbol{r}_{h}\right)+H_{\mathrm{e}}\left(\boldsymbol{p}_{\mathrm{e}}\right)+H_{\mathrm{h}}\left(\boldsymbol{p}_{\mathrm{\mathrm{h}}}\right)\label{eq:Hpeph}
\end{equation}
with the energy $E_{\mathrm{g}}$ of the band gap and the
Coulomb interaction, which is screened by the 
dielectric constant~$\varepsilon$:
\begin{equation}
V\left(\boldsymbol{r}_{e}-\boldsymbol{r}_{h}\right)=-\frac{e^{2}}{4\pi\varepsilon_{0}\varepsilon}\frac{1}{\left|\boldsymbol{r}_{e}-\boldsymbol{r}_{h}\right|}.
\end{equation}
The exchange interaction as well as the central-cell corrections described in Refs.~\cite{1,100} 
are not included in the Hamiltonian~(\ref{eq:Hpeph}) as they do not 
affect the exciton states treated in Secs.~\ref{sec:Experimental} and~\ref{sub:field}.

In the presence of an external magnetic field, the
corresponding Hamiltonian is obtained via the minimal substitution. 
After introducing relative and center of mass coordinates~\cite{90,91}
and setting the position and momentum of the 
center of mass to zero, the complete Hamiltonian
of the relative motion reads~\cite{34,33,39,TOE,44,90,91}
\begin{eqnarray}
H & = & E_{\mathrm{g}}+V\left(\boldsymbol{r}\right)+H_{B}\nonumber \\
 & + & H_{\mathrm{e}}\left(\boldsymbol{p}+e\boldsymbol{A}\left(\boldsymbol{r}\right)\right)+H_{\mathrm{h}}\left(-\boldsymbol{p}+e\boldsymbol{A}\left(\boldsymbol{r}\right)\right).\label{eq:H}
\end{eqnarray}
Since the magnetic field $\boldsymbol{B}$ is constant in our experiments, we use the vector potential $\boldsymbol{A}=\left(\boldsymbol{B}\times\boldsymbol{r}\right)/2$.
The term $H_{B}$ describes the energy
of the spins in the magnetic field~\cite{25,44,44_12,33}:
\begin{equation}
H_{B}=\mu_{\mathrm{B}}\left[g_{c}\boldsymbol{S}_{\mathrm{e}}+\left(3\kappa+g_{s}/2\right)\boldsymbol{I}-g_{s}\boldsymbol{S}_{\mathrm{h}}\right]\cdot\boldsymbol{B}/\hbar.
\end{equation}
Here $\mu_{B}$ denotes the Bohr magneton, $g_{s}\approx2$ the $g$-factor
of the hole spin $S_{\mathrm{h}}$, $g_{c}$ the $g$-factor of the
conduction band or the electron spin $S_{\mathrm{e}}$.
The value of the fourth Luttinger parameter $\kappa$
is unknown and will be determined in Sec.~\ref{sub:field}. 
In the case of a finite spin-orbit coupling~$\Delta$
an additional term would generally appear in $H_{B}$, which depends on
the fifth Luttinger parameter $q$~\cite{25,44,44_12}. However, this
term is connected with spin-orbit interactions of higher order~\cite{44}
and is not considered here.
All material values used in our calculations are listed in Table~\ref{tab:1}.

\begin{table}

\protect\caption{Material parameters used in the calculations. 
Note that the value of $\gamma_{1}'$
is a result of the analysis in Sec.~\ref{sub:field}
and slightly differs from the value $\gamma_{1}'=2.77$ of Ref.~\cite{80}.\label{tab:1}}

\begin{centering}
\begin{tabular}{lll}
\hline 
band gap energy & $E_{\mathrm{g}}=2.17208\,\mathrm{eV}$ & \cite{GRE}\tabularnewline
electron mass & $m_{\mathrm{e}}=0.99\, m_{0}$ & \cite{M2}\tabularnewline
dielectric constant & $\varepsilon=7.5$ & \cite{SOK1_82L1}\tabularnewline
spin-orbit coupling & $\Delta=0.131\,\mathrm{eV}$ & \cite{80}\tabularnewline
valence band parameters & $\gamma_{1}'=2.74$ & \cite{80}\tabularnewline
 & $\mu'=0.0586$ & \cite{80}\tabularnewline
 & $\delta'=-0.404$ & \cite{80}\tabularnewline
 & $\eta_1=-0.02$ & \cite{80}\tabularnewline 
 & $\nu=2.167$ & \cite{80}\tabularnewline
 & $\tau=1.5$ & \cite{80}\tabularnewline
$g$-factor of cond. band & $g_{\mathrm{c}}=2.1$ & \cite{29a}\tabularnewline
\hline 
\end{tabular}
\par\end{centering}

\end{table}

For the case that the magnetic field is oriented along one of the
directions of high symmetry, i.e., along the $[001]$, $[110]$
or $[111]$ direction,
we rotate the coordinate system to make
the quantization axis coincide with the direction of the magnetic field
and then express the Hamiltonian~(\ref{eq:H}) in terms of irreducible tensors~\cite{ED,7_11,44} (see
Appendix~\ref{sub:Hamiltonianrm}).
We can then calculate a matrix representation of
the Schr\"odinger equation corresponding to the Hamiltonian (\ref{eq:H})
using a complete basis. 

As regards the angular momentum part of the basis, we have to
consider that the different parts of the Hamiltonian couple the
quasispin $I$, the hole spin $S_{\mathrm{h}}$, and the angular momentum $L$
of the exciton. Due to the spin orbit coupling $H_{\mathrm{so}}$ and the cubic part of the
Hamiltonian~(\ref{eq:H}), we introduce the
effective hole spin $J=I+S_{\mathrm{h}}$ and the angular momentum $F=L+J$.
We finally include the electron spin in our basis by introducing the
total angular momentum $F_{t}=F+S_{\mathrm{e}}$.
For the radial
part of the exciton wave function we use the Coulomb-Sturmian functions~\cite{S1}
\begin{equation}
U_{NL}\left(r\right)=N_{NL}\left(2\rho\right)^{L}e^{-\rho}L_{N}^{2L+1}\left(2\rho\right)\label{eq:U}
\end{equation}
with $\rho=r/\alpha$, a normalization factor $N_{NL}$,
the associated Laguerre polynomials $L_{n}^{m}\left(x\right)$ and
an arbitrary scaling parameter $\alpha$. The radial quantum number
$N$ is related to the principal quantum number $n$ via $n=N+L+1$.
Finally, we use the following ansatz for the exciton wave function
\begin{subequations}
\begin{eqnarray}
\left|\Psi\right\rangle  & = & \sum_{NLJFF_{t}M_{F_{t}}}c_{NLJFF_{t}M_{F_{t}}}\left|\Pi\right\rangle,\\
\nonumber \\
\left|\Pi\right\rangle  & = & \left|N,\, L;\,\left(I,\, S_{\mathrm{h}}\right)\, J;\, F,\, S_{\mathrm{e}};\, F_{t},\, M_{F_{t}}\right\rangle\label{eq:basis}
\end{eqnarray}\label{eq:ansatz}%
\end{subequations}
with real coefficients $c$. The parenthesis and semicolons in
Eq.~(\ref{eq:basis}) shall illustrate the coupling scheme of the
spins and the angular momenta.

Inserting the ansatz~(\ref{eq:ansatz}) in the Schr\"odinger
equation $H\Psi=E\Psi$ and multiplying from the left with
another basis state $\left\langle \Pi'\right|$,
we obtain a matrix representation 
of the Schr\"odinger equation of the form
\begin{equation}
\boldsymbol{D}\boldsymbol{c}=E\boldsymbol{M}\boldsymbol{c}.\label{eq:gev}
\end{equation}
The vector $\boldsymbol{c}$ contains the coefficients of the ansatz~(\ref{eq:ansatz}).
All matrix elements, which enter the symmetric matrices $\boldsymbol{D}$ and
$\boldsymbol{M}$ and which have not been treated in Ref.~\cite{100},
are given in the Appendices~\ref{sub:Matrix-elements} and~\ref{sub:Reduced-matrix-elements}.
The generalized eigenvalue problem~(\ref{eq:gev})
is finally solved using an appropriate LAPACK routine~\cite{Lapack}.

\section{Oscillator strengths \label{sec:Oscillator}}

With the solutions of the eigenvalue problem~(\ref{eq:gev}) one can directly
calculate the relative oscillator strengths for the transitions 
from the ground state of the solid to the exciton states.
In doing so, four important points need to be considered~\cite{100}:

(i) The transition is parity forbidden, for which reason the transition matrix element
is proportional to the gradient of the envelope function at $r=0$ and
the exciton state must have a component with angular momentum $L=1$. 
(ii) The dipole operator does not change the total spin $S=S_{\mathrm{e}}+S_{\mathrm{h}}=0$
of the electron and the hole. 
(iii) The total symmetry of the exciton state must 
be identical to the symmetry of the dipole operator.
(iv) The quasispin $I$ transforms according to $\Gamma_5^+$ whereas
a normal spin one transforms according to $\Gamma_4^+$. Therefore, since
$\Gamma_5^+=\Gamma_2^+ \otimes\Gamma_4^+$ holds for the cubic group~\cite{G1},
one has to multiply all symmetries with $\Gamma_2^+$~\cite{28}.

The dipole operator transforms according to the 
irreducible representation $D^1$ of the full rotation 
group. Hence, it transforms in $\mathrm{Cu_{2}O}$ according to the irreducible representation $\Gamma_{4}^{-}$
of the cubic group $O_{\mathrm{h}}$. 
Since the coupling of $S=0$, $L=1$ and $I=1$ leads to a total
angular momentum of $F_t=0$, $F_t=1$ and $F_t=2$, one has to find in this nine-dimensional space the
three correct linear combinations of states, which transform according to
$\Gamma_{4}^{-}$.
The reduction of the representations $D^{F_t}$ of the full rotation group with 
the inclusion of the additional factor $\Gamma_2^+$ yields~\cite{G3}
\begin{subequations}
\begin{align}
\tilde{D}^{0}=D^{0}\otimes\Gamma_{2}^{+} &\: =\Gamma_{1}^{-}\otimes\Gamma_{2}^{+}=\Gamma_{2}^{-},\\
\displaybreak[1]
\nonumber \\
\tilde{D}^{1}=D^{1}\otimes\Gamma_{2}^{+} &\: =\Gamma_{4}^{-}\otimes\Gamma_{2}^{+}=\Gamma_{5}^{-},\\
\displaybreak[1]
\nonumber \\
\tilde{D}^{2}=D^{2}\otimes\Gamma_{2}^{+} &\: =\left(\Gamma_{3}^{-}\oplus\Gamma_{5}^{-}\right)\otimes\Gamma_{2}^{+}\nonumber \\
 &\: =\Gamma_{3}^{-}\oplus\Gamma_{4}^{-}.
\end{align}
\end{subequations}
We see that $\Gamma_{4}^{-}$ is included only in the reduction of 
the irreducible representation $D^{F_t}$ with $F_t=2$.
Using the relations given in Ref.~\cite{G1}, one can finally find the three linear combinations
of $F_t=2$ states, which transform according to $\Gamma_{4}^{-}$ and which are connected to light being
linearly polarized in $x$, $y$, and $z$ direction:
\begin{subequations}
\begin{align}
\left|\pi_x\right\rangle= &\; \frac{i}{\sqrt{2}}\left[\left|2,\,-1\right\rangle_D+\left|2,\,1\right\rangle_D\right],\\
\left|\pi_y\right\rangle= &\; \frac{1}{\sqrt{2}}\left[\left|2,\,-1\right\rangle_D-\left|2,\,1\right\rangle_D\right],\\
\left|\pi_z\right\rangle= &\; \frac{i}{\sqrt{2}}\left[\left|2,\,-2\right\rangle_D-\left|2,\,2\right\rangle_D\right].
\end{align}
\label{eq:Dxyz}%
\end{subequations}
Here $\left|F_t,\,M_{F_t}\right\rangle_D$ is a short notation for the state
\begin{eqnarray}
& & \left|\left(S_{\mathrm{e}},\,S_{\mathrm{h}}\right)\,S,\,I;\,I+S,\,L;\,F_t,\,M_{F_t}\right\rangle\nonumber\\
& = & \left|\left(1/2,\,1/2\right)\,0,\,1;\,1,\,1;\,F_t,\,M_{F_t}\right\rangle,
\end{eqnarray}
in which the coupling scheme of the spins and angular momenta is different from the one of Eq.~(\ref{eq:basis})
\begin{equation}
S_{\mathrm{e}}+S_{\mathrm{h}}=S\quad\rightarrow\quad(I+S)+L=F_t
\end{equation}
due to the requirement that $S$ must be a good quantum number.

To determine the relative oscillator strength of an arbitrary exciton state with 
the wave function $\Psi$ [see Eq.~(\ref{eq:ansatz})], we have to account 
for its $\left|\pi_x\right\rangle$, $\left|\pi_y\right\rangle$ or $\left|\pi_z\right\rangle$ component.
Hence, the relative oscillator strength for light being linearly polarized in $x$ direction is, e.g., given by 
\begin{equation}
f_{\mathrm{rel}}\sim\left|\lim_{r\rightarrow0}\frac{\partial}{\partial r}\left\langle \pi_x\middle|\Psi\left(\boldsymbol{r}\right)\right\rangle\right|^2\label{eq:frel}
\end{equation}
with $r=\left|\boldsymbol{r}\right|$ (see also Appendix~\ref{sub:Oscillator-strengths}).
Note that when considering the coupling scheme of the angular momenta and spins in Eq.~(\ref{eq:basis}) there are only three combinations
of $L=1$, $J$, and $F$, which lead to a total angular momentum of $F_t=2$. These are
\begin{subequations}
\begin{eqnarray}
L=1,\,J=1/2\quad & \rightarrow & \quad F=3/2,\\
L=1,\,J=3/2\quad & \rightarrow & \quad F=3/2,\\
L=1,\,J=3/2\quad & \rightarrow & \quad F=5/2.
\end{eqnarray}
\end{subequations}
Hence, the oscillator strength is definitely zero if all the coefficients $c_{N1\frac{1}{2}\frac{3}{2}2M_{F_{t}}}$, $c_{N1\frac{3}{2}\frac{3}{2}2M_{F_{t}}}$, and $c_{N1\frac{3}{2}\frac{5}{2}2M_{F_{t}}}$ in the exciton state $\Psi$ are zero [cf. Eq.~(\ref{eq:ansatz})].


In the presence of an external magnetic field the operator $\boldsymbol{A}_{\mathrm{rad}}\boldsymbol{p}$,
which describes the interaction between the radiation field or light and the exciton and which enters
the dipole matrix element,
has to be replaced by $\boldsymbol{A}_{\mathrm{rad}}\left[\boldsymbol{p}+e\boldsymbol{A}\left(\boldsymbol{r}\right)\right]$
due to the minimal substitution.
However, the second term is generally small in comparison to the first one 
and it vanishes in the Faraday configuration considered here~\cite{FT,PAE,PAE_E17}.

Nevertheless, we have to consider that the magnetic field reduces the 
symmetry of the system. Furthermore, since the incident light is oriented parallel to $\boldsymbol{B}$
and since we choose the quantization axis parallel to $\boldsymbol{B}$,
we have to find the correct linear combinations of the $F_t=2$ states, which describe 
linearly or circularly polarized light
for the three orientations of the magnetic field considered here.
This will be done in the following.

\subsection{Magnetic field in [001] direction \label{sub:001}}

In a magnetic field, which is oriented along the $[001]$ direction, the
symmetry $O_{\mathrm{h}}$ of the system is reduced to $C_{\mathrm{4h}}$ 
and we have to consider the reduction of the 
irreducible representation $\Gamma_{4}^{-}$ of $O_{\mathrm{h}}$ by the group $C_{\mathrm{4h}}$:
\begin{equation}
\Gamma_{4}^{-} \rightarrow \Gamma_{1}^{-}\oplus\Gamma_{3}^{-}\oplus\Gamma_{4}^{-}.
\end{equation}
Using the method of projection operators~\cite{G1},
we can determine the correct linear combinations of the states
in Eq.~(\ref{eq:Dxyz}) which transform according to the irreducible
representations of $C_{\mathrm{4h}}$:
\begin{subequations}
\begin{alignat}{3}
\Gamma_{1}^{-}: &\; |\pi_z\rangle && = \frac{i}{\sqrt{2}}\left[\left|2,\,-2\right\rangle_D-\left|2,\,2\right\rangle_D\right],\\
\nonumber \\
\Gamma_{3}^{-}: &\; |\sigma_z^+\rangle && =\frac{-i}{\sqrt{2}}\left[\left|\pi_x\right\rangle+i\left|\pi_y\right\rangle\right] = \left|2,\,-1\right\rangle_D,\\
\nonumber \\
\Gamma_{4}^{-}: &\; |\sigma_z^-\rangle && =\frac{i}{\sqrt{2}}\left[\left|\pi_x\right\rangle-i\left|\pi_y\right\rangle\right] = -\left|2,\,1\right\rangle_D.
\end{alignat}
\label{eq:Dpi001}%
\end{subequations}
One can see that $\Gamma_{1}^{-}$ is connected with light 
which is linearly polarized along $[001]$, i.e., in the $z$ direction. 
This light cannot be observed along the $z$ axis
due to the Faraday configuration in the experiment.
$\Gamma_{3}^{-}$ as well as $\Gamma_{4}^{-}$ are connected with circularly
polarized light. Consequently, only states of the symmetry $\Gamma_{3}^{-}$
or $\Gamma_{4}^{-}$ can be observed in absorption experiments and we calculate the 
relative oscillator strengths by evaluating
\begin{equation}
f_{\mathrm{rel}}\sim\left|\lim_{r\rightarrow0}\frac{\partial}{\partial r}\left\langle \sigma_z^{\pm}\middle|\Psi\left(\boldsymbol{r}\right)\right\rangle\right|^2.
\end{equation}
Note that the sign of $\sigma^{\pm}$ is defined by the direction of rotation of the polarisation
with respect to $\boldsymbol{B}$. 

\subsection{Magnetic field in [110] direction \label{sub:110}}

In a magnetic field, which is directed in $[110]$ direction, the
symmetry $O_{\mathrm{h}}$ of the system is reduced to $C_{\mathrm{2h}}$.
In this case the reduction of the irreducible representation 
$\Gamma_{4}^{-}$ of $O_{\mathrm{h}}$ by the group $C_{\mathrm{4h}}$ reads
\begin{equation}
\Gamma_{4}^{-} \rightarrow \Gamma_{1}^{-}\oplus\Gamma_{2}^{-}\oplus\Gamma_{2}^{-}
\end{equation}
and the correct linear combinations of the states
in Eq.~(\ref{eq:Dxyz}) which transform according to the irreducible
representations of $C_{\mathrm{2h}}$ are
\begin{subequations}
\begin{alignat}{2}
\Gamma_{1}^{-}: &\; \frac{1}{\sqrt{2}}\left[\left|\pi_x\right\rangle+\left|\pi_y\right\rangle\right],\\
\nonumber\\
\Gamma_{2}^{-}: &\; \left|\pi_z\right\rangle,\\
\nonumber\\
\Gamma_{2}^{-}: &\; \frac{1}{\sqrt{2}}\left[\left|\pi_x\right\rangle-\left|\pi_y\right\rangle\right].
\end{alignat}
\label{eq:Dpi110}%
\end{subequations}
We see that $\Gamma_{1}^{-}$ is connected with light which is linearly polarized along $[110]$
and that $\Gamma_{2}^{-}$ is connected with transverse polarized light in $[001]$ and $[1\bar{1}0]$ direction.
Since the states
$|\pi_x\rangle$ and $|\pi_y\rangle$ transform according to the
same irreducible representation, we can also use the
following linear combinations, which describe circularly polarized light:
\begin{equation}
\Gamma_{2}^{-}: \mp\frac{i}{\sqrt{2}}\left[\left|\pi_z\right\rangle\pm \frac{i}{\sqrt{2}}\left[\left|\pi_x\right\rangle-\left|\pi_y\right\rangle\right]\right].
\end{equation}

We now choose the quantization axis parallel to $\boldsymbol{B}$, i.e.,
we rotate the coordinate system by the Euler angles
$\left(\alpha,\,\beta,\,\gamma\right)=\left(\pi,\,\pi/2,\,\pi/4\right)$.
This coordinate transformation reads
\begin{equation}
\boldsymbol{r}'=\left(\begin{array}{c}
x'\\
y'\\
z'
\end{array}\right)=\frac{1}{\sqrt{2}}\left(\begin{array}{ccc}
0 & 0 & \sqrt{2}\\
1 & -1 & 0\\
1 & 1 & 0
\end{array}\right)\left(\begin{array}{c}
x\\
y\\
z
\end{array}\right)=\boldsymbol{R}\boldsymbol{r}.
\end{equation}
Note that the direction of the $x'$ and $y'$ axis are predefined by the crystal axes.
Rotating the states of Eq.~(\ref{eq:Dpi110}) as well yields
\begin{subequations}
\begin{alignat}{3}
\Gamma_{1}^{-}: &\; |\pi_{z'}\rangle && = \frac{1}{\sqrt{2}}\left[\left|2,\,-1\right\rangle_D-\left|2,\,1\right\rangle_D\right],\\
\nonumber\\
\Gamma_{2}^{-}: &\; |\pi_{x'}\rangle && = \frac{\sqrt{3}}{2}\left|2,\,0\right\rangle_D\nonumber\\
& && +\frac{1}{\sqrt{8}}\left[\left|2,\,-2\right\rangle_D+\left|2,\,2\right\rangle_D\right],\\
\nonumber\\
\Gamma_{2}^{-}: &\; |\pi_{y'}\rangle && = \frac{i}{\sqrt{2}}\left[\left|2,\,2\right\rangle_D-\left|2,\,-2\right\rangle_D\right],
\end{alignat}
\label{eq:Dpi110r}%
\end{subequations}
or
\begin{subequations}
\begin{alignat}{4}
\Gamma_{2}^{-}: &\; |\sigma_{z'}^{\pm}\rangle && = \mp \frac{i}{\sqrt{2}}\left[|\pi_{x'}\rangle\pm i|\pi_{y'}\rangle\right].
\end{alignat}
\label{eq:Dpi110s}%
\end{subequations}

Finally, we calculate the 
relative oscillator strengths by evaluating
\begin{equation}
f_{\mathrm{rel}}\sim\left|\lim_{r\rightarrow0}\frac{\partial}{\partial r}\left\langle \pi_{x',y'}\middle|\Psi\left(\boldsymbol{r}\right)\right\rangle\right|^2\label{eq:frel}
\end{equation}
for light which is polarized in [001] or in $[1\bar{1}0]$ direction or
\begin{equation}
f_{\mathrm{rel}}\sim\left|\lim_{r\rightarrow0}\frac{\partial}{\partial r}\left\langle \sigma_{z'}^{\pm}\middle|\Psi\left(\boldsymbol{r}\right)\right\rangle\right|^2
\end{equation}
for circularly polarized light.


\subsection{Magnetic field in [111] direction \label{sub:111}}

In a magnetic field, which is directed in $[111]$ direction, the
symmetry $O_{\mathrm{h}}$ of the system is reduced to $C_{\mathrm{3i}}$.
In this case we have
\begin{equation}
\Gamma_{4}^{-} \rightarrow \Gamma_{1}^{-}\oplus\Gamma_{2}^{-}\oplus\Gamma_{3}^{-}
\end{equation}
and the correct linear combinations of the states
in Eq.~(\ref{eq:Dxyz}) read
\begin{subequations}
\begin{alignat}{2}
\Gamma_{1}^{-}: &\; \frac{1}{\sqrt{3}}\left[\left|\pi_x\right\rangle+\left|\pi_y\right\rangle+\left|\pi_z\right\rangle\right],\\
\nonumber\\
\Gamma_{2}^{-}: &\; \frac{-i}{\sqrt{2}}\left[\frac{1}{\sqrt{6}}\left[\left|\pi_x\right\rangle+\left|\pi_y\right\rangle-2\left|\pi_z\right\rangle\right]\right.\nonumber\\
& \qquad+\left.i\frac{1}{\sqrt{2}}\left[-\left|\pi_x\right\rangle+\left|\pi_y\right\rangle\right]\right],\\
\nonumber\\
\Gamma_{3}^{-}: &\; \frac{i}{\sqrt{2}}\left[\frac{1}{\sqrt{6}}\left[\left|\pi_x\right\rangle+\left|\pi_y\right\rangle-2\left|\pi_z\right\rangle\right]\right.\nonumber\\
& \qquad-\left.i\frac{1}{\sqrt{2}}\left[-\left|\pi_x\right\rangle+\left|\pi_y\right\rangle\right]\right].
\end{alignat}
\label{eq:Dpi111}%
\end{subequations}
We see that $\Gamma_{1}^{-}$ is connected with light which is linearly polarized along [111]
and that $\Gamma_{2}^{-}$ and $\Gamma_{3}^{-}$ are connected with circularly polarized light transverse to this axis.

For $\boldsymbol{B}\parallel [111]$ we rotate the coordinate system by the Euler angles
$\left(\alpha,\,\beta,\,\gamma\right)=\left(0,\,\arccos(1/\sqrt{3}),\,\pi/4\right)$.
This coordinate transformation reads
\begin{equation}
\boldsymbol{r}''=\left(\begin{array}{c}
x''\\
y''\\
z''
\end{array}\right)=\frac{1}{\sqrt{6}}\left(\begin{array}{ccc}
1 & 1 & -2\\
-\sqrt{3} & \sqrt{3} & 0\\
\sqrt{2} & \sqrt{2} & \sqrt{2}
\end{array}\right)\left(\begin{array}{c}
x\\
y\\
z
\end{array}\right)=\boldsymbol{R}\boldsymbol{r}.
\end{equation}
Rotating the states of Eq.~(\ref{eq:Dpi111}) yields
\begin{subequations}
\begin{alignat}{3}
\Gamma_{1}^{-}: &\; |\pi_{z''}\rangle && =\left|2,\,0\right\rangle_D,\\
\nonumber\\
\Gamma_{2}^{-}: &\; |\sigma_{z''}^+\rangle && = \frac{i}{\sqrt{3}}\left[\sqrt{2}\left|2,\,-2\right\rangle_D-\left|2,\,1\right\rangle_D\right],\\
\nonumber\\
\Gamma_{3}^{-}: &\; |\sigma_{z''}^-\rangle && = \frac{-i}{\sqrt{3}}\left[\sqrt{2}\left|2,\,2\right\rangle_D+\left|2,\,-1\right\rangle_D\right],
\end{alignat}
\label{eq:Dpi111r}%
\end{subequations}
and we calculate the 
relative oscillator strengths by evaluating
\begin{equation}
f_{\mathrm{rel}}\sim\left|\lim_{r\rightarrow0}\frac{\partial}{\partial r}\left\langle \sigma_{z''}^{\pm}\middle|\Psi\left(\boldsymbol{r}\right)\right\rangle\right|^2.
\end{equation}

\section{Experiment \label{sec:Experimental}}

In the experiment, we investigated the absorption $\alpha$ of $\mathrm{Cu_{2}O}$
crystal slabs that were cut and polished from a natural rock.
Three different samples with different orientations were prepared:
In the first sample the $[001]$ direction is normal to the crystal 
surface, in the other two samples the normal direction corresponds to the 
$[110]$ and $[111]$ orientation, respectively. The thicknesses of these 
samples varied from $30$ to $50\,\mathrm{\upmu m}$ which is, however, of 
no relevance for the results described below. 
For application of a magnetic field, the samples were inserted in a 
superconducting split coil magnet with a helium cryostat in the center, 
in which the samples were cooled down to $T=1.4\,\mathrm{K}$. The magnetic 
field direction was chosen to be along the optical axis (Faraday-configuration), i.e., 
also normal to the studied crystal slabs. 

\begin{figure}
\begin{centering}
\includegraphics[width=0.95\columnwidth]{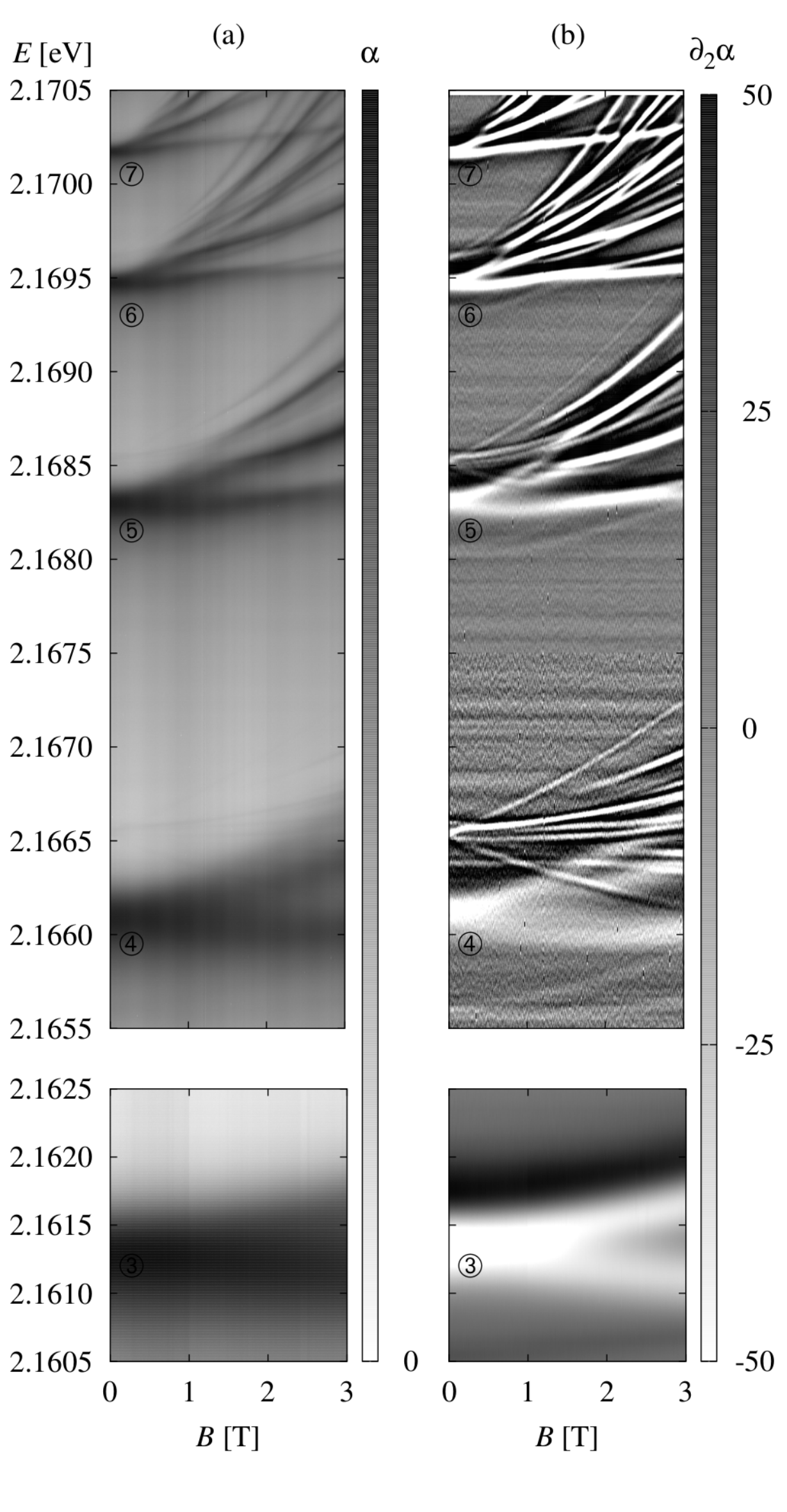}
\par\end{centering}

\protect\caption{(a) Absorption spectra of the $n=3\ldots 7$ excitons in an external magnetic field $\boldsymbol{B}\parallel [001]$ 
composed by superposition of counter-circularly polarized spectra. The absorption constant $\alpha$ is given in arbitrary units. 
(b) Second derivative
of the experimental absorption. The value of $\partial_2\alpha$ is given in arbitrary units.
The second derivative levels intensity differences and thus highlights the contribution of
exciton states with higher angular momentum $L$ to the spectra.
Note that the faint periodic horizontal patterns are artefacts arising 
from taking the second deviative of the spectra.~\label{fig:abs_app}}

\end{figure}

The absorption was measured using a white light source which was filtered 
by a double monochromator such that only the energy range in which the 
exciton states of interest are located was covered. 
The exciting light was circularly polarized by a quarter-wave retarder 
and was sent under normal incidence onto the sample with a spot size of 
about $100\,\mathrm{\upmu m}$.
The transmitted light was dispersed by another double monochromator 
and detected by a liquid-nitrogen cooled charge coupled device camera, 
providing a spectral resolution of about $10\,\mathrm{\upmu eV}$. Since 
the spectral width of the studied exciton resonances is significantly 
larger than this resolution, the setup provides sufficient resolution, as 
confirmed also by reference measurements with a tunable frequency-stabilized 
laser with neV resolution. The measurements with the two different light sources 
yielded identical spectra. 
The excitation density was chosen low enough that the excitation of dressed 
states as discussed in P.~Gr\"unwald \emph{et~al.}~\cite{76}, can be neglected 
and the observed spectral lines in the experiment correspond to resonant absorption peaks.

For an overview, we give in Fig.~\ref{fig:abs_app}
a contour plot of absorption spectra composed by superposition of counter-circularly 
polarized spectra to show all optically accessible exciton states, recorded on the 
[001] oriented sample. The energy range from the $n=3$ exciton around $2.162\,\mathrm{eV}$ up 
to $n=7$ on the high energy side is displayed. The left hand plot shows the absorption as recorded, 
and the right hand side shows the second derivative of the recorded spectra which helps 
to level intensity differences between features and highlight 
weak absorption lines. At zero field the spectra are dominated by the dipole-allowed 
$P$ excitons~\cite{28,100}. With increasing magnetic field, each $P$ exciton shows 
mostly a doublet splitting. Simultaneously on their high energy flanks new lines 
emerge and become steadily stronger. Except of the magnitude of the splitting, 
the appearance of the $n=4$ multiplet is somewhat similar to that of the $n=5$ multiplet, 
involving also a similar number of lines.
By contrast, the $n=6$ multiplet is composed of many more lines 
due to the presence of exciton states with angular momentum $L=5$.

\begin{figure}
\begin{centering}
\includegraphics[width=1.0\columnwidth]{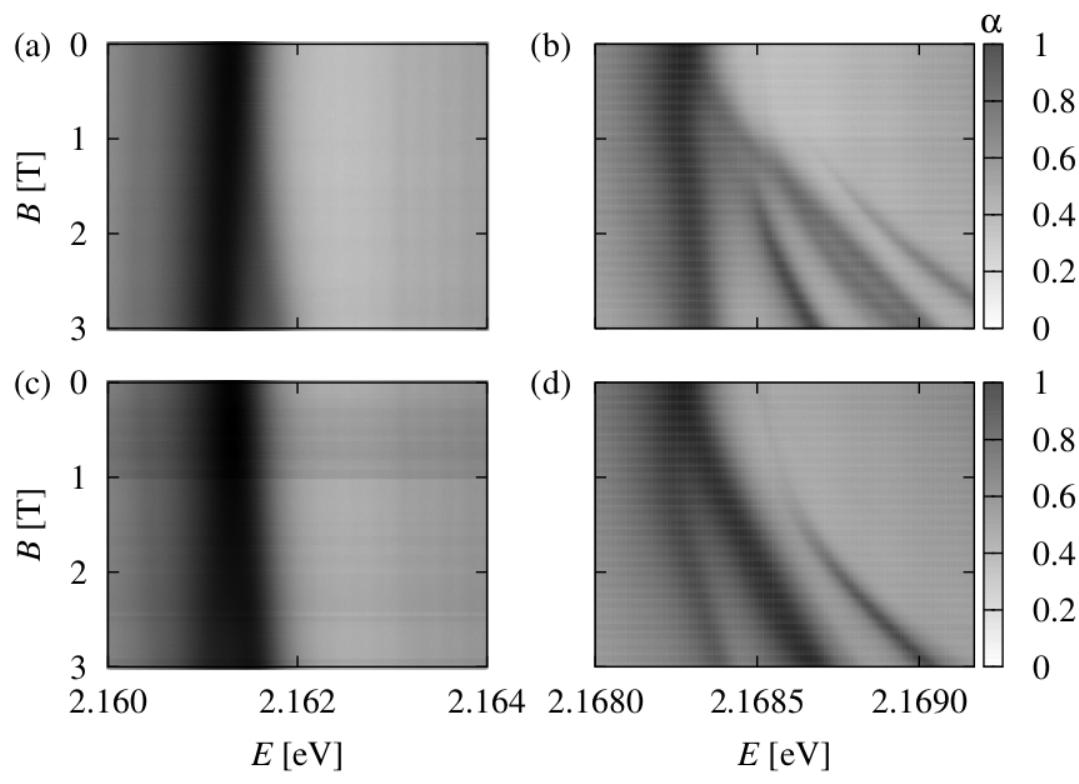}
\par\end{centering}

\protect\caption{Absorption spectra of (left) $n=3$ and (right) $n=5$ excitons.
The values of the absorption constant $\alpha$ are given in arbitrary units.
The spectra were obtained in Faraday configuration with 
an external magnetic field $\boldsymbol{B}\parallel [001]$ and
(above) $\sigma^{+}$ or (below) $\sigma^{-}$ polarized light.~\label{fig:exp}}
\end{figure}

\begin{figure}
\begin{centering}
\includegraphics[width=1.0\columnwidth]{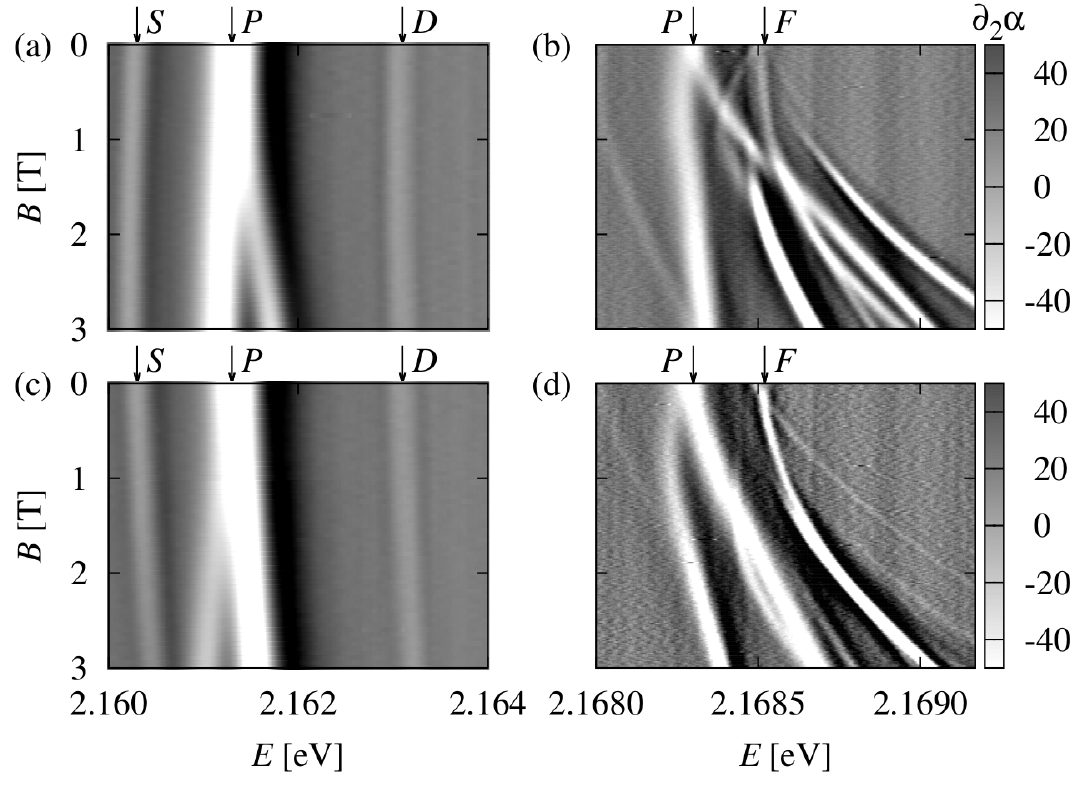}
\par\end{centering}

\protect\caption{Second derivative $\left(\partial_2\alpha\right)$ of the absorption spectra of Fig.~\ref{fig:exp}. 
The values of $\partial_2\alpha$ are given in arbitrary units.
The large number of exciton lines
for $n=5$ and
especially the zero-field splitting of $P$ and $F$ excitons
can be seen more clearly in this presentation of the experimental results.
We note that for the $n=3$ exciton two more lines are observed in the respresentation, 
which we attribute to the $S$ exciton for the low energy line and the $D$ exciton for 
the high energy line. We attribute their appearance to quadrupole-allowed transitions 
to the $S$ exciton. For the $D$ exciton quadrupole excitation is possible because of 
mixing with the $S$ exciton~\cite{7}.~\label{fig:exp1}}
\end{figure}

\begin{figure}
\begin{centering}
\includegraphics[width=1.0\columnwidth]{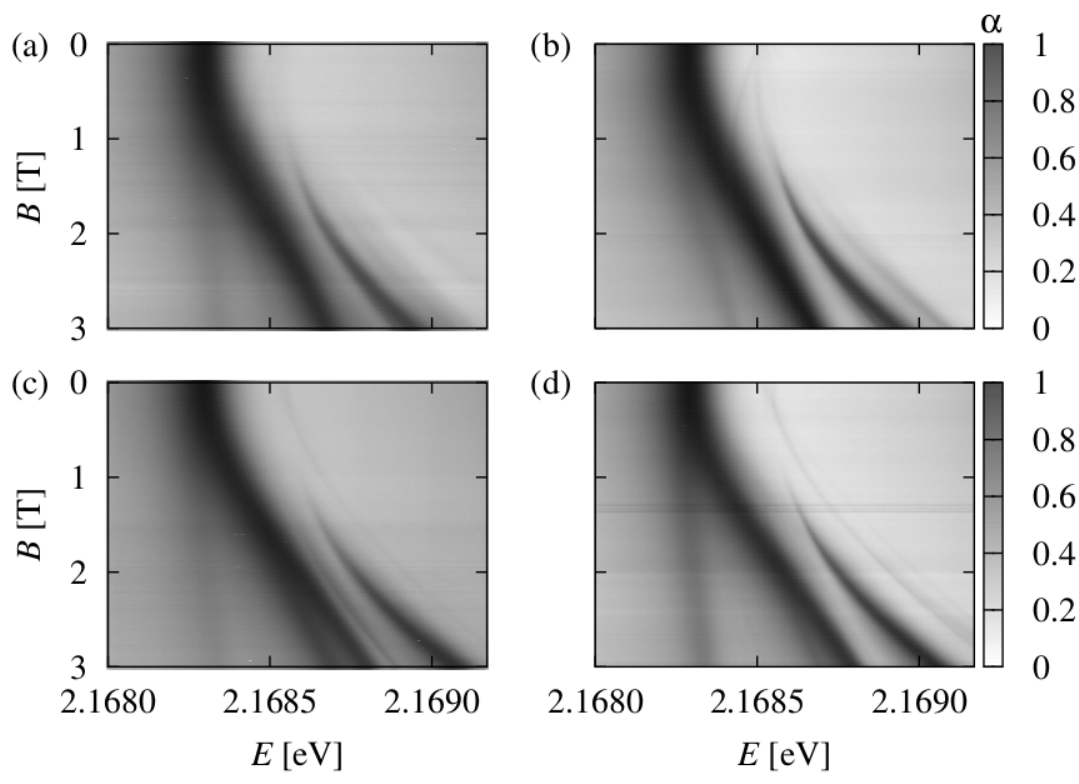}
\par\end{centering}

\protect\caption{Absorption spectra of
the $n=5$ exciton for (left) $\boldsymbol{B}\parallel [110]$ and (right) $\boldsymbol{B}\parallel [111]$.
The values of $\alpha$ are given in arbitrary units.
Also for these orientations of the magnetic field, one can see a clear difference between the spectra
for (above) $\sigma^{+}$ and (below) $\sigma^{-}$ polarized light.~\label{fig:exp2}}
\end{figure}

\begin{figure}
\begin{centering}
\includegraphics[width=1.0\columnwidth]{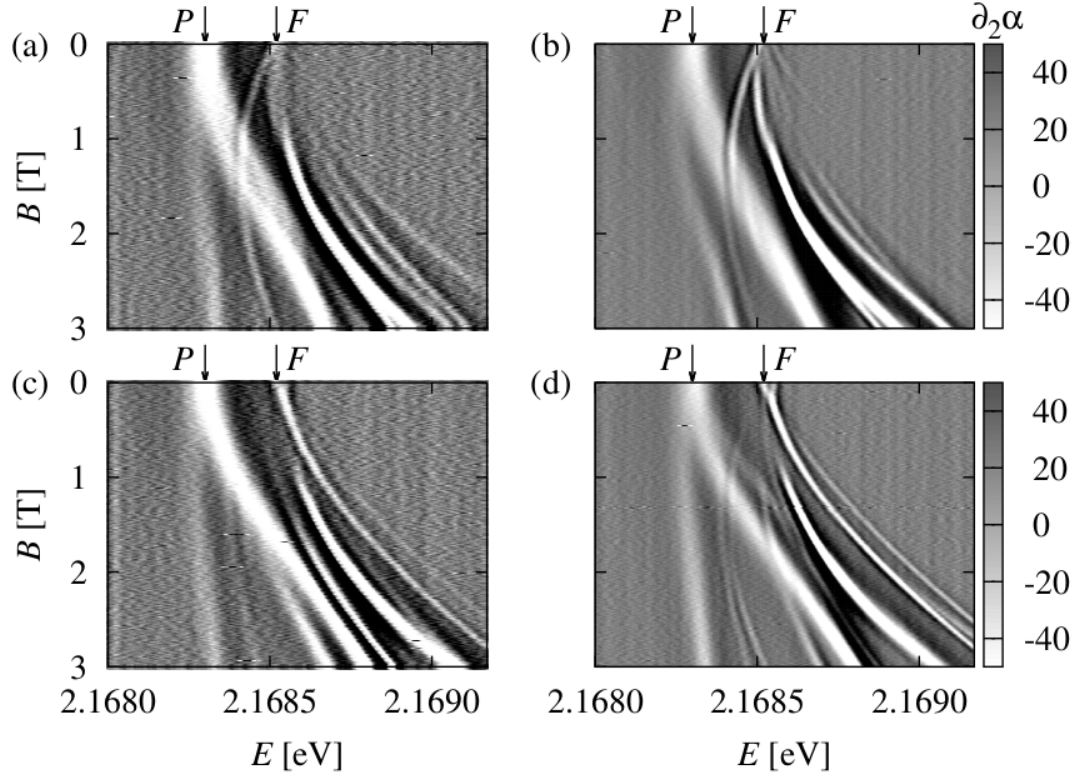}
\par\end{centering}

\protect\caption{Second derivative of the absorption spectra of Fig.~\ref{fig:exp2}.
The values of $\partial_2\alpha$ are given in arbitrary units.
Again, the large number of exciton lines
for $n=5$ can be seen more clearly
in this presentation of the experimental results.~\label{fig:exp3}}
\end{figure}

In general, the impact of the magnetic field in terms of shifting
and splitting levels grows with increasing
principal quantum number $n$ due to the increased
extension of the exciton wave function, 
\begin{equation}
r_{n,L} = \frac{1}{2} a_{\mathrm{exc}} \left( 3 n^2 - L(L+1) \right)
\end{equation}
with the exciton Bohr radius $a_{\mathrm{exc}} = 1.11\,\mathrm{nm}$ and envelope angular 
momentum $L$, compared to the magnetic length 
\begin{equation}
\ell_c = \sqrt{\frac{\hbar}{eB}} = \frac{25.6\,\mathrm{nm}}{\sqrt{B\,\mathrm{[T]}}}.
\end{equation}
For example, the diamagnetic shift of the center-of-gravity of a line
multiplet belonging to a specific $n$, which is a measure
of the wave function extension normal to the field, increases
from less than $0.1\,\mathrm{meV}$ for $n=3$ up to $B=3\,\mathrm{T}$ (see below) 
to about $0.5\,\mathrm{meV}$ for $n=5$. Furthermore, due to the reduced energy 
splitting between exciton multiplets and their extended Zeeman-splitting in a magnetic 
field for higher $n$, exciton states belonging to different principal quantum numbers 
come into resonance at smaller magnetic field strengths. For example, the first resonance 
of two states belonging to $n=5$ and $n=6$ occurs at about $3.5\,\mathrm{T}$, while the corresponding 
resonance between the $n=6$ and the $n=7$ multiplet occurs at $2\,\mathrm{T}$.

We have shown recently that for $n>6$ the electron-hole motion becomes chaotic in a magnetic field, 
as confirmed by corresponding theoretical calculations~\cite{QC,175}. In the chaos regime the density 
of states is so high, that an exact identification of the individual exciton states, while 
still being distinguishable, becomes increasingly complex as does the theoretical calculation. 
Instead statistical methods can be applied, such as the calculation of the level spacing distribution 
which transfers from a Poissonian to a Wigner-Dyson distribution going from the regular to the chaotic 
motion regime. The Wigner-Dyson distribution function is characterized by the dominance of avoided 
crossings between levels while crossings are suppressed. For $n\leq 6$ we observe in our measurements 
clearly a dominance of crossings, confirming that in this range the motion stays regular. In combination 
with the possibility to assign the exciton character, we therefore restrict to $n\leq 5$ here.

To obtain more detailed insight into the level splitting in a magnetic field, 
we focus on exciton multiplets belonging to a particular $n$. 
Moreover, we consider circularly polarized spectra with the goal to 
reduce the number of exciton lines in a spectrum.
Fig.~\ref{fig:exp} shows contour plots of absorption spectra for 
$\boldsymbol{B}\parallel [001]$ in the energy range of the 
$n=3$ and $n=5$ excitons, in one panel for left circular 
polarization of the white light, and in the other panel 
for right circular polarization. Again, also the second derivatives 
of the absorption spectra are shown (see Fig.~\ref{fig:exp1}). 
The $n=3$ exciton, for which the envelope 
angular momentum $L$ is limited to $2$, is characterized by the doublet splitting 
of the $P$ exciton. The faint features from the $S$ and $D$ excitons due to 
quadrupole-allowed transitions also show a doublet splitting. We note that 
the relative oscillator strengths to a good approximation do not change 
with magnetic field, which indicates that as expected these excitons do not 
become mixed with the $P$ excitons. For completeness we stress again that in a 
crystal orbital angular momentum  strictly speaking is no good quantum number, 
but the discrete symmetry leads to a mixing of states with different $L$, in 
particular of the $S$ and the $D$ excitons as well as the $P$ and the $F$ excitons. 
Still, for reasons of simplicity, we use these notations here.

The magnetic field induces and enhances the mixing of states with the same parity. 
This is clearly seen for the $n=5$ exciton multiplet. At $B=0\,\mathrm{T}$, 
besides the dominant $P$ exciton the $F$ excitons splitting to higher energies can be 
resolved, most prominently in the second derivative spectra due to their rather 
small relative oscillator strength (more than two orders of magnitude smaller than that 
of the $P$ states). With increasing field the $F$ excitons become much more prominent 
and also new lines from this multiplet emerge due to state mixing with the $P$ excitons, 
from which oscillator strength is transferred. This allows us to obtain a detailed 
picture of the Zeeman-effect induced splitting of the different lines. Comparing the 
spectra for the two counter-circular polarizations we note that they show distinct 
differences. This clearly shows that the often used description of an exciton by a 
hydrogen atom-model is not appropriate, but the details of the electronic band structure 
need to be accounted for.

This is corroborated by the measurements in Figs.~\ref{fig:exp2} and~\ref{fig:exp3}, showing corresponding 
measurements and second derivatives for the $n=5$ exciton in the 
crystals with [110] and [111] orientation. Also here we find striking 
differences between the two counter-circular polarizations. Moreover, 
they are also different compared to the spectra for the [001] field orientation. 
Also indications for anticrossings together
with the corresponding exchange of oscillator strength 
can be observed in the field dispersion, see, for example, the $\sigma^-$-polarized 
spectrum in the lower right panel of Fig.~\ref{fig:exp3}.

\begin{figure}
\begin{centering}
\includegraphics[width=1.0\columnwidth]{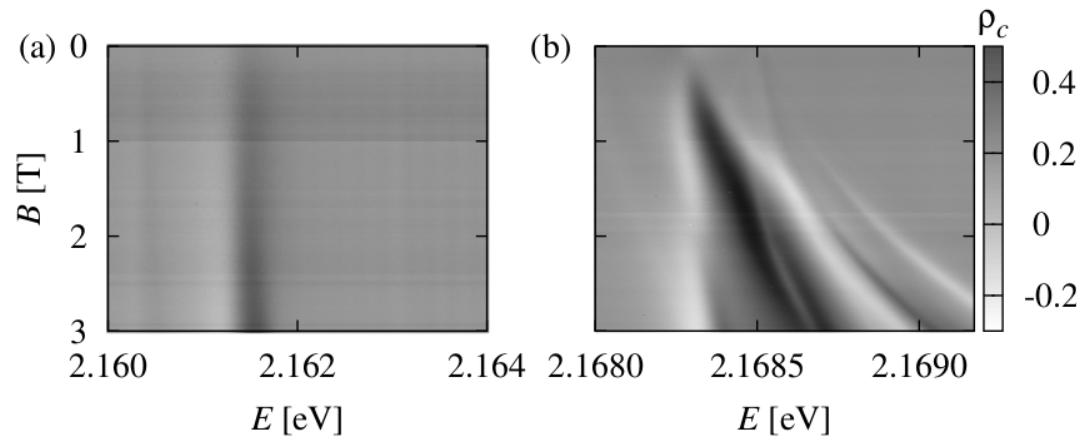}
\par\end{centering}

\protect\caption{Degree of circular polarization $\rho_c$ of (a) $n=3$ and (b) $n=5$ excitons
for $\boldsymbol{B}\parallel [001]$.
It can be seen that none of the observed lines shows complete circular polarization.
~\label{fig:exp4}}
\end{figure}

We note that for the [001] orientation a perfect circular polarization 
of the transitions is expected, so that excitation with particular 
helicity in principle can be used for optical orientation of the electron spin.
However, when calculating the degree of circular polarization defined as
\begin{eqnarray}
\rho_c \left ( E \right) = \frac{ \alpha_{\sigma^-} - \alpha_{\sigma^+}}{ \alpha_{\sigma^-} + \alpha_{\sigma^+}},
\end{eqnarray}
where $\alpha_{\sigma^-}$  and $\alpha_{\sigma^+}$ are the absorption as 
function of energy for $\sigma^-$ and $\sigma^+$ polarization, respectively, 
we find polarization degrees well below unity but they are limited to values 
between $-0.3<\rho_c <0.5$, as seen from Fig.~\ref{fig:exp4}. 
This reduced polarization is result of the finite width of the lines leading 
to spectral overlap. Note, that for the other field orientation such as along 
[110] the polarization of the optical transitions is expected to be more complex.

\section{Results and discussion\label{sub:field}}

First we determine the maximum number of dipole-allowed exciton states
via group theoretical considerations.
In this way, we can also show that
the external magnetic field lifts all degeneracies of the exciton states.
In the spherical approximation, in which the cubic
part of the Hamiltonian is neglected, the angular momentum $F$ is a good quantum
number. However, if the complete Hamiltonian is treated, the reduction
of the irreducible representations $D^{F}$ of the rotation group
by the cubic group $O_{\mathrm{h}}$ has to be considered~\cite{G1}. 
As has already been stated in Sec.~\ref{sec:Oscillator},
a normal spin one transforms according to the irreducible representation
$\Gamma_{4}^{+}$ of the cubic group and the quasispin $I$ transforms
according to $\Gamma_{5}^{+}=\Gamma_{4}^{+}\otimes\Gamma_{2}^{+}$. Therefore,
one has to include the additional factor $\Gamma_{2}^{+}$ when determining
the symmetry of an exciton state~\cite{7,28,100}. This symmetry is given by the symmetry
of the envelope function, the valence band, and the conduction
band:
\begin{equation}
\Gamma_{\mathrm{exc}}=\Gamma_{\mathrm{env}}\otimes\Gamma_{\mathrm{v}}\otimes\Gamma_{\mathrm{c}}.
\end{equation}



As the quasispin $I$ already enters the angular momentum $F$, we obtain the 
combined symmetry of the envelope function and the hole in the reduction of
the representations $D^{F}$ of the full rotation group:
\begin{subequations}
\begin{align}
\tilde{D}^{\frac{1}{2}}=D^{\frac{1}{2}}\otimes\Gamma_{2}^{+} &\: =\Gamma_{6}^{-}\otimes\Gamma_{2}^{+}=\Gamma_{7}^{-},\label{eq:12-1}\\
\displaybreak[1]
\nonumber \\
\tilde{D}^{\frac{3}{2}}=D^{\frac{3}{2}}\otimes\Gamma_{2}^{+} &\: =\Gamma_{8}^{-}\otimes\Gamma_{2}^{+}=\Gamma_{8}^{-},\\
\displaybreak[1]
\nonumber \\
\tilde{D}^{\frac{5}{2}}=D^{\frac{5}{2}}\otimes\Gamma_{2}^{+} &\: =\left(\Gamma_{7}^{-}\oplus\Gamma_{8}^{-}\right)\otimes\Gamma_{2}^{+}\nonumber \\
 &\: =\Gamma_{6}^{-}\oplus\Gamma_{8}^{-},\label{eq:52-1}\\
\displaybreak[1]
\nonumber \\
\tilde{D}^{\frac{7}{2}}=D^{\frac{7}{2}}\otimes\Gamma_{2}^{+} &\: =\left(\Gamma_{6}^{-}\oplus\Gamma_{7}^{-}\oplus\Gamma_{8}^{-}\right)\otimes\Gamma_{2}^{+}\nonumber \\
 &\: =\Gamma_{7}^{-}\oplus\Gamma_{6}^{-}\oplus\Gamma_{8}^{-}.\label{eq:72-1}
\end{align}
\end{subequations}
It can be seen that there are two exciton states for $n=2$ or $n=3$ and
seven exciton states for $n=4$ or $n=5$. Including the symmetry $\Gamma_{6}^{+}$
of the electron or the conduction band, the total symmetry of the exciton
is
\begin{subequations}
\begin{align}
\Gamma_{\mathrm{exc}}^{\frac{1}{2}}=\tilde{D}^{\frac{1}{2}}\otimes\Gamma_{6}^{+} &\: =\left(\Gamma_{2}^{-}\oplus\Gamma_{5}^{-}\right),\label{eq:g12}\\
\displaybreak[1]
\nonumber \\
\Gamma_{\mathrm{exc}}^{\frac{3}{2}}=\tilde{D}^{\frac{3}{2}}\otimes\Gamma_{6}^{+} &\: =\left(\Gamma_{3}^{-}\oplus\Gamma_{4}^{-}\oplus\Gamma_{5}^{-}\right),\\
\displaybreak[1]
\nonumber \\
\Gamma_{\mathrm{exc}}^{\frac{5}{2}}=\tilde{D}^{\frac{5}{2}}\otimes\Gamma_{6}^{+} &\: =\left(\Gamma_{1}^{-}\oplus\Gamma_{4}^{-}\right)\nonumber \\
 &\: \oplus\left(\Gamma_{3}^{-}\oplus\Gamma_{4}^{-}\oplus\Gamma_{5}^{-}\right),\\
\displaybreak[1]
\nonumber \\
\Gamma_{\mathrm{exc}}^{\frac{7}{2}}=\tilde{D}^{\frac{7}{2}}\otimes\Gamma_{6}^{+} &\: =\left(\Gamma_{1}^{-}\oplus\Gamma_{4}^{-}\right)\nonumber \\
 &\: \oplus\left(\Gamma_{2}^{-}\oplus\Gamma_{5}^{-}\right)\nonumber \\
 &\: \oplus\left(\Gamma_{3}^{-}\oplus\Gamma_{4}^{-}\oplus\Gamma_{5}^{-}\right),\label{eq:g72}
\end{align}
\label{eq:Gexc}%
\end{subequations}
respectively. Since the symmetries in parentheses belong to degenerate
states, there are fourfold and eightfold degenerate states.

\begin{figure}
\begin{centering}
\includegraphics[width=0.95\columnwidth]{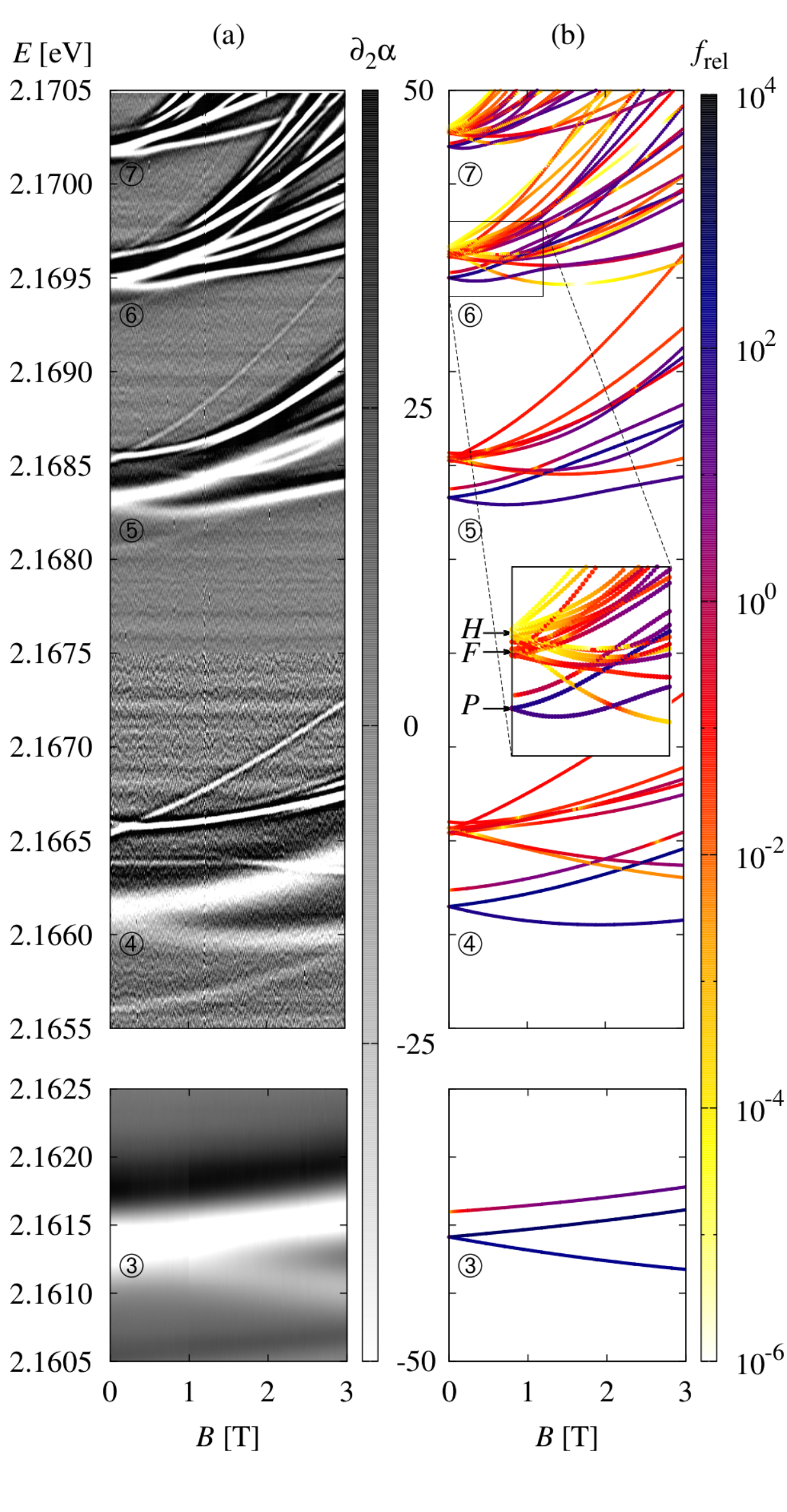}
\par\end{centering}

\protect\caption{Spectra of the $n=3\ldots 7$ excitons in an external magnetic field $\boldsymbol{B}\parallel [001]$
for $\sigma^{-}$ polarized light. (a) Second derivative
of the experimental absorption spectrum. The value of $\partial_2\alpha$ is given in arbitrary units.
(c) Theoretical line spectrum for $\kappa=-0.5$. The colorbar shows the calculated
relative oscillator strength in arbitrary units.
Since the complete Hamiltonian mixes exciton states with odd angular momentum,
we expect the appearance of $F$ excitons for $n\geq 4$ and $H$ excitons for $n\geq 6$ as
can be clearly seen, e.g., in the inset in panel (b).~\label{fig:abs_theor}}

\end{figure}

\begin{figure}
\begin{centering}
\includegraphics[width=1.0\columnwidth]{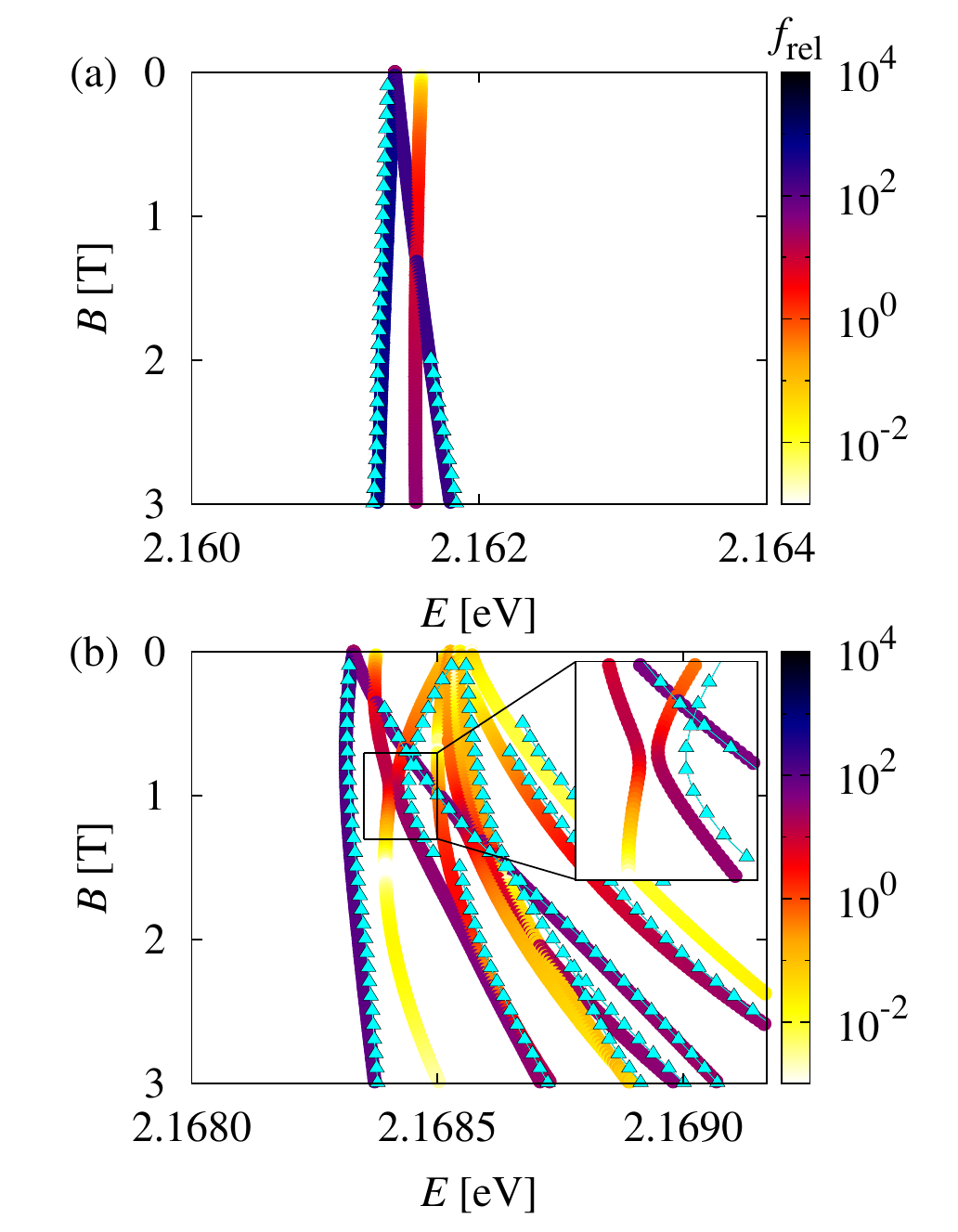}
\par\end{centering}

\protect\caption{Theoretical line spectrum of the
(a) $n=3$ and (b) $n=5$ exciton states
in an external magnetic field $\boldsymbol{B}\parallel [001]$ for $\sigma^{+}$ polarized light.
The colorbar shows the calculated
relative oscillator strength in arbitrary units.
The inset enlarges 
the most prominent anticrossing in the spectrum. 
This anticrossing involves the
two exciton states, which originate from the 
$\Gamma_7^{-}$ state of Eq.~(\ref{eq:12-1}) and the $\Gamma_7^{-}$ state of 
Eq.~(\ref{eq:72-1}) at $B=0\,\mathrm{T}$ (see also Ref.~\cite{100}).
By comparing the theoretical results 
with the position of those exciton states, which could unambiguously
be read out from the experimental spectrum (blue triangles) 
using the method of the second derivative~\cite{80},
we can determine the fourth Luttinger parameter~$\kappa$.
An excellent agreement between theory and experiment is obtained
for $\kappa=-0.50\pm 0.10$.
As the second derivative does not yield the exact position of the resonances,
we have shifted the experimental spectrum by an amount of (a) $100~\mathrm{\upmu eV}$
and (b) $55~\mathrm{\upmu eV}$.~\label{fig:Line-spectrum}}
\end{figure}

\begin{figure}
\begin{centering}
\includegraphics[width=1.0\columnwidth]{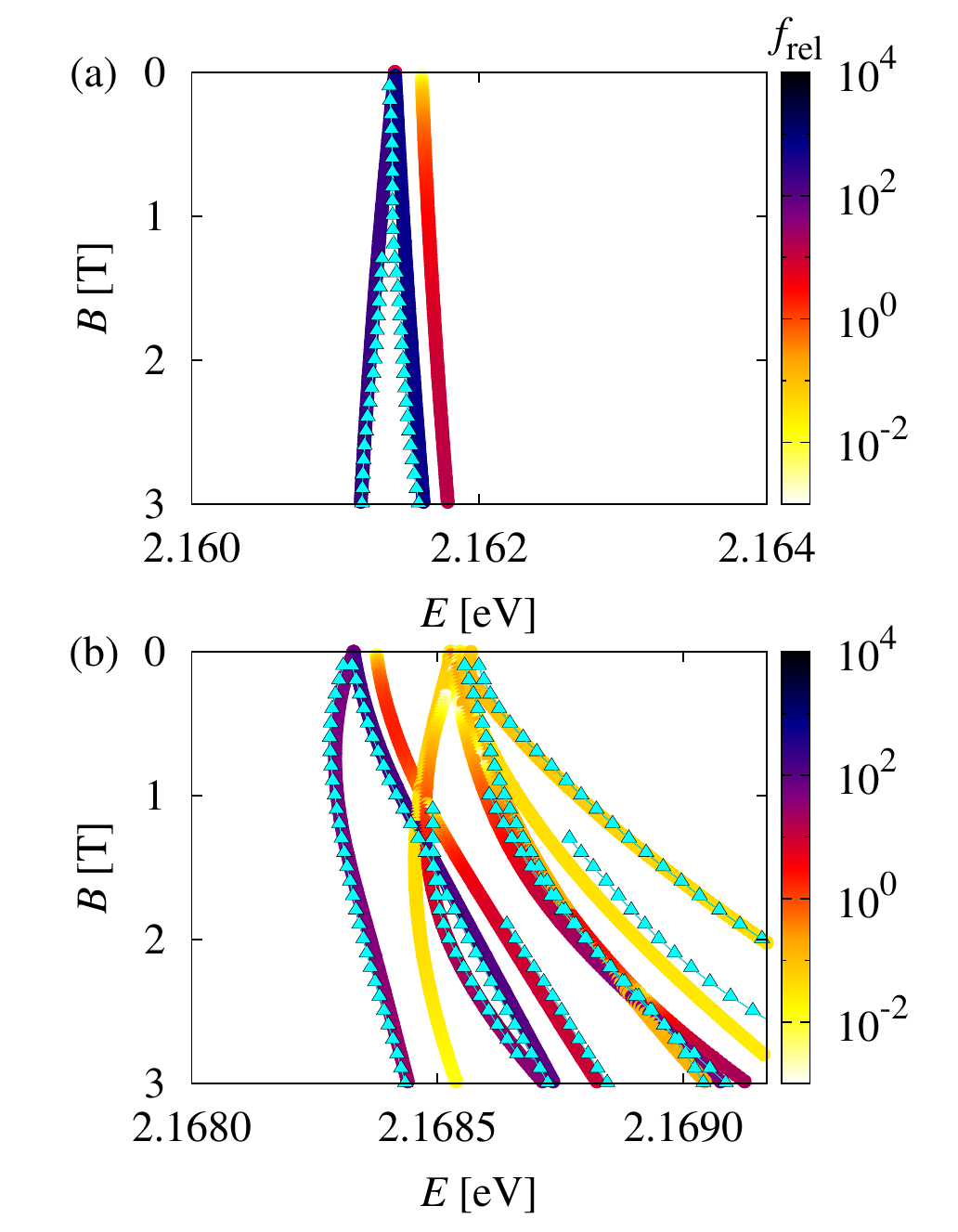}
\par\end{centering}

\protect\caption{
Same comparison as in Fig.~\ref{fig:Line-spectrum} but for 
$\sigma^{-}$ polarized light. 
Again, we obtain an excellent agreement
between theory and experiment for $\kappa=-0.50\pm 0.10$.
~\label{fig:Line-spectrum2}}
\end{figure}

\begin{figure}
\begin{centering}
\includegraphics[width=1.0\columnwidth]{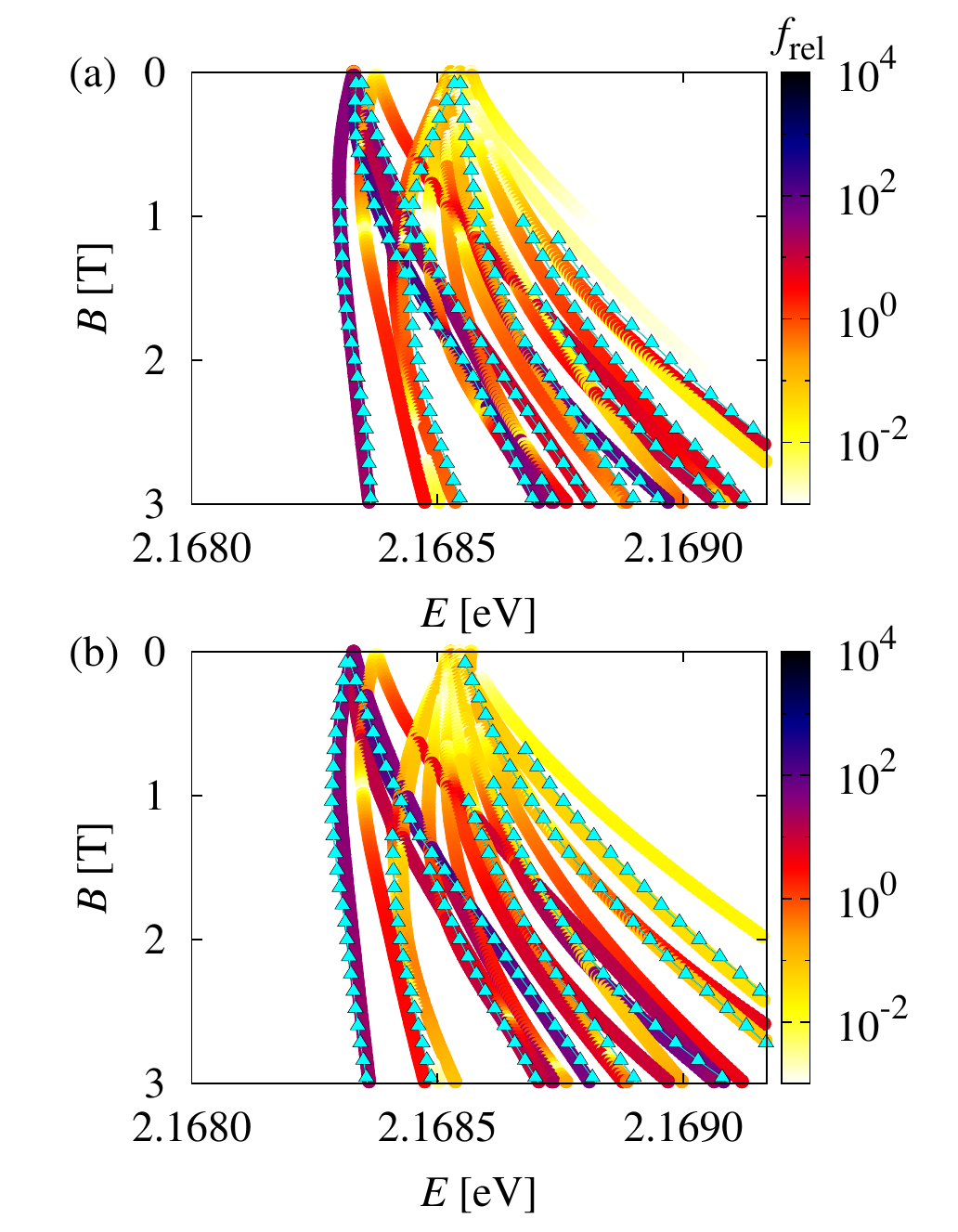}
\par\end{centering}

\protect\caption{Theoretical line spectrum of the
$n=5$ exciton states
in an external magnetic field $\boldsymbol{B}\parallel [110]$ for (a) $\sigma^{+}$ and (b) $\sigma^{-}$ polarized light.
The colorbar shows the calculated
relative oscillator strength in arbitrary units.
The exciton states, which could unambiguously
be read out from the experiment are again marked by blue triangles.
Since $\sigma^{+}$ and $\sigma^{-}$ polarized light belong to the 
same irreducible representation of $C_{\mathrm{2h}}$,
it is possible to excite a certain exciton state
by $\sigma^+$ \emph{and} by $\sigma^-$ polarized light. Hence, all 20 dipole-allowed 
exciton states can be observed in panel~(a) and in panel~(b).
Note that we have shifted the experimental spectrum by an amount of $26~\mathrm{\upmu eV}$.~\label{fig:Line-spectrum3}}
\end{figure}

\begin{figure}
\begin{centering}
\includegraphics[width=1.0\columnwidth]{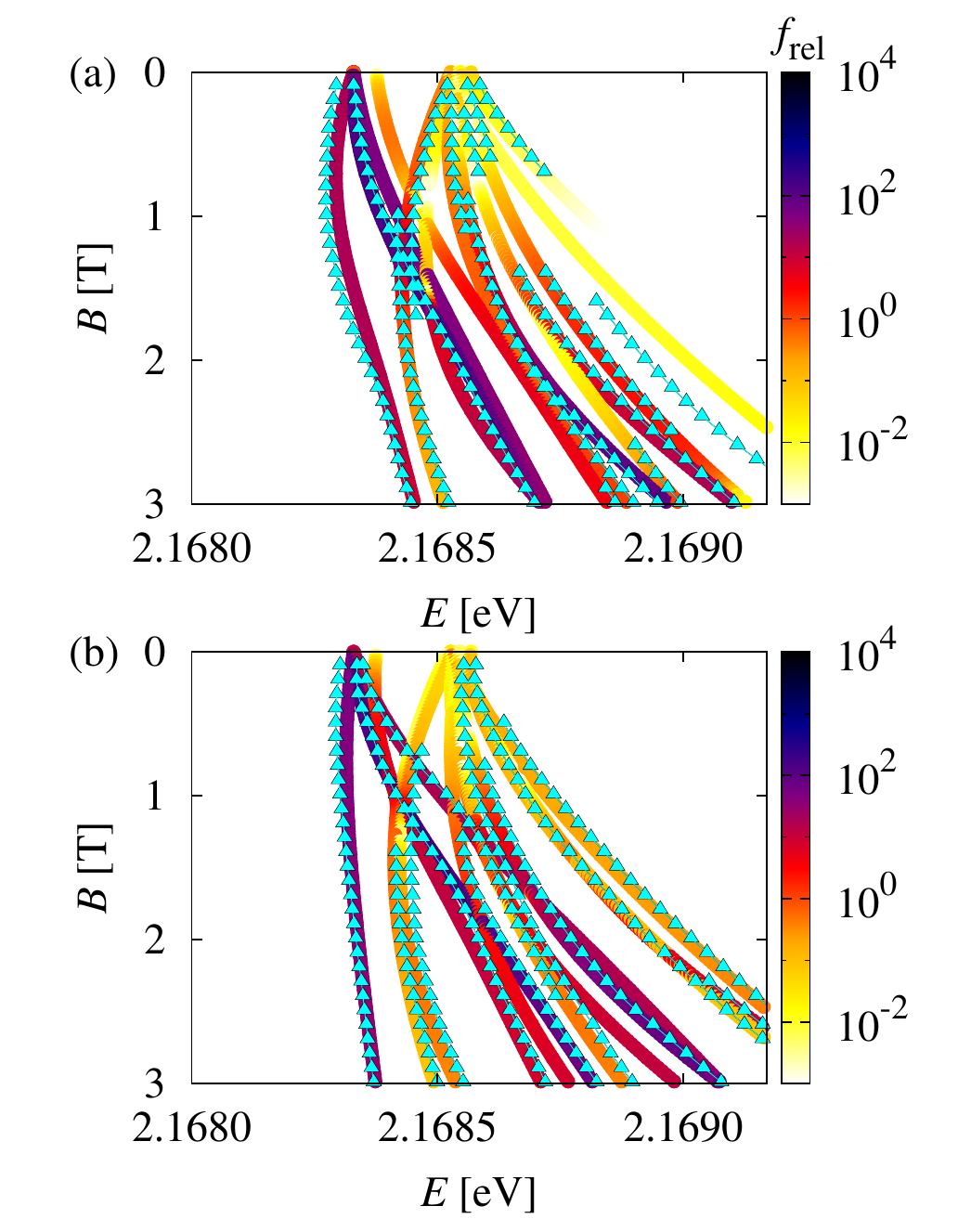}
\par\end{centering}

\protect\caption{Theoretical line spectrum of the
$n=5$ exciton states
in an external magnetic field $\boldsymbol{B}\parallel [111]$ for (a) $\sigma^{+}$ and (b) $\sigma^{-}$ polarized light.
Note that we have shifted the experimental spectrum by an amount of $26~\mathrm{\upmu eV}$.~\label{fig:Line-spectrum4}}
\end{figure}

In the presence of a magnetic field being oriented
along one of the crystal axis, we have to consider the 
reduction of the irreducible representations
of the cubic group $O_{\mathrm{h}}$ by the group $C_{\mathrm{4h}}$~\cite{G3}:
\begin{subequations}
\begin{align}
\Gamma_{1}^{-} \rightarrow &\: \Gamma_{1}^{-},\\
\Gamma_{2}^{-} \rightarrow &\: \Gamma_{2}^{-},\\
\Gamma_{3}^{-} \rightarrow &\: \Gamma_{1}^{-}\oplus\Gamma_{2}^{-},\\
\Gamma_{4}^{-} \rightarrow &\: \Gamma_{1}^{-}\oplus\Gamma_{3}^{-}\oplus\Gamma_{4}^{-},\label{eq:C4h4}\\
\Gamma_{5}^{-} \rightarrow &\: \Gamma_{2}^{-}\oplus\Gamma_{3}^{-}\oplus\Gamma_{4}^{-}.
\end{align}
\label{eq:C4h}%
\end{subequations}
Inserting these relations in Eq.~(\ref{eq:Gexc}), we see that
degeneracies are further lifted. Since all irreducible representations
of $C_{\mathrm{4h}}$ are one-dimensional, we expect that all degeneracies
are lifted so that there are 12 exciton states for $n=2$ or $n=3$
and 40 exciton states for $n=4$ or $n=5$. Since only the states with the symmetry $\Gamma_3^-$
and $\Gamma_4^-$ are dipole-allowed, we immediately see that out of these only
$6$ states are dipole-allowed for $n=2$ or $n=3$
and $20$ for $n=4$ or $n=5$.

In the same manner we can also treat the special
cases of a magnetic field being oriented along the 
$[110]$ and the $[111]$ direction, i.e.,
we have to consider the 
reduction of the irreducible representations
of the cubic group $O_{\mathrm{h}}$ by the groups $C_{\mathrm{2h}}$
and $C_{\mathrm{3i}}$~\cite{G3}.
Since all irreducible representations
of these groups are one-dimensional we also expect that all degeneracies
are lifted and that there are 12 exciton states for $n=2$ or $n=3$
and 40 exciton states for $n=4$ or $n=5$.
However, if we consider only the number of dipole-allowed states we find that
there are $6$ $\left(n=2,\,3\right)$ and $20$ $\left(n=4,\,5\right)$
dipole-allowed states for $\boldsymbol{B}\parallel [110]$ but
$8$ $\left(n=2,\,3\right)$ and $26$ $\left(n=4,\,5\right)$
dipole-allowed states for $\boldsymbol{B}\parallel [111]$.

\begin{figure}
\begin{centering}
\includegraphics[width=1.0\columnwidth]{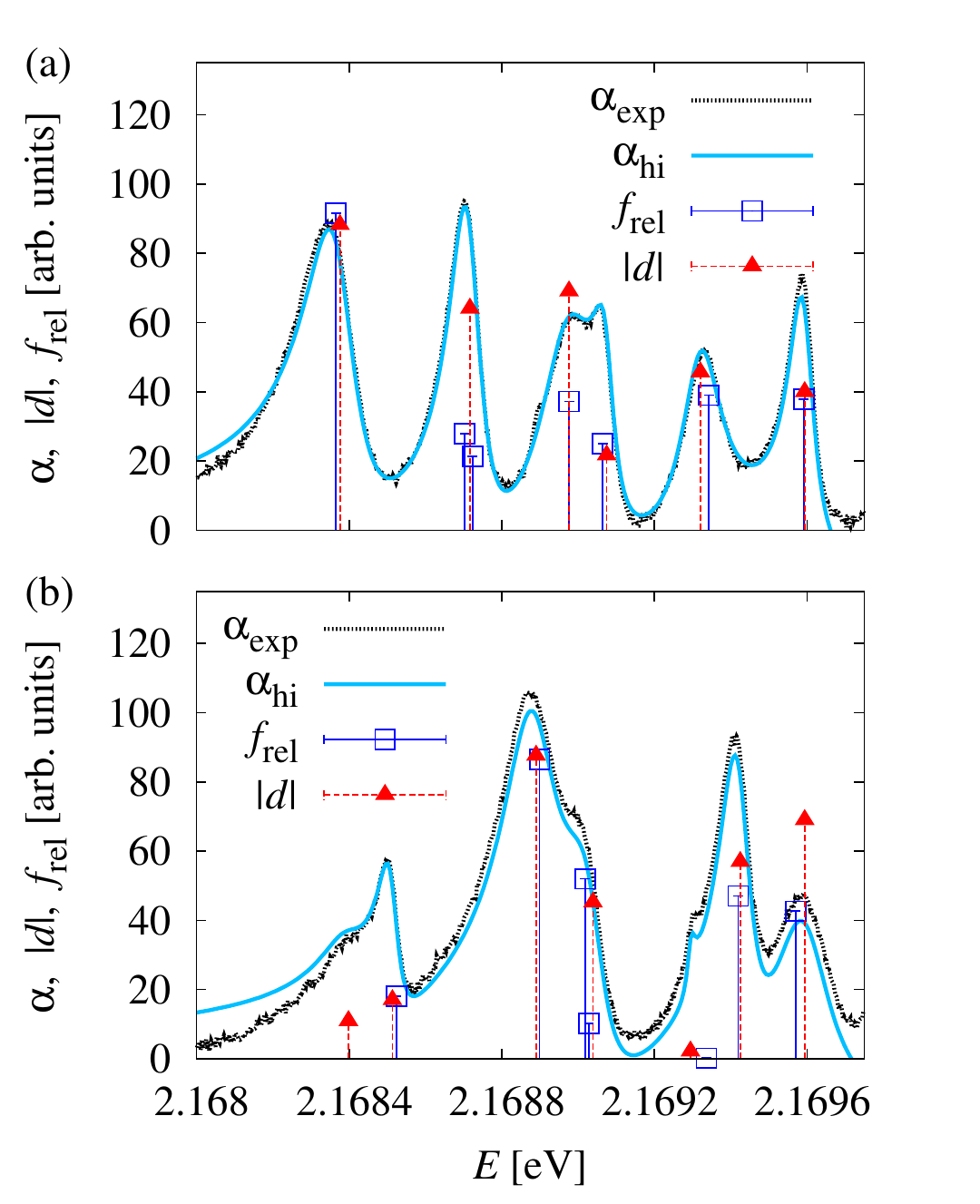}
\par\end{centering}

\protect\caption{The experimental spectrum (black dotted line) for $\boldsymbol{B}\parallel [001]$
is analyzed using the method of harmonic inversion~\cite{HI1,HI2}.
The panels show the spectra for (a) $\sigma^{+}$ polarized light at $B=3\,\mathrm{T}$ and 
(b) $\sigma^{-}$ polarized light at $B=4\,\mathrm{T}$.
Since several exciton states are almost degenerate for 
$\sigma^{-}$ polarized light at $B=3\,\mathrm{T}$ (cf.~Fig.~\ref{fig:Line-spectrum2}),
we analyze the spectrum at $B=4\,\mathrm{T}$.
The positions and the amplitudes of the resonances
obtained are marked by red triangles.
The blue solid line shows the function $G\left(E\right)$ of Eq.~(\ref{eq:Gw})
for these resonances.
Note that in panel (a) the rightmost resonance originates 
from an $n=6$ exciton state.
Comparing the positions of the resonances to the theoretical
spectrum (dark blue squares)
yields the optimum values $\gamma_1=1.73\pm0.02$ and $\kappa=-0.50\pm 0.10$ 
for the first and the fourth Luttinger parameter.
Furthermore, we obtain a good agreement between the 
relative oscillator strengths $f_\mathrm{rel}$
and the modulus $\left|d\right|$ of amplitudes 
of the resonances.~\label{fig:hi}}

\end{figure}

For a first overview we show the experimental spectra for $\boldsymbol{B}\parallel [001]$ along with
theoretical results for $n=3\ldots 7$ in Fig.~\ref{fig:abs_theor}.
In the special case of $\gamma_2=\gamma_3=\eta_i=0$ the angular momentum
$L$ would be a good quantum number and due to the selection rules 
discussed in Sec.~\ref{sec:Oscillator}
only $P$ excitons would be observable in this case. However, even without a
magnetic field the complete Hamiltonian couples different exciton states with
odd values of $L$ so that also exciton states with higher 
angular momentum gain a small oscillator strength.
This can be seen clearly from the 
theoretically calculated spectrum in the panel (b) of Fig.~\ref{fig:abs_theor}.

Since $L\leq n-1$ holds, the number of exciton lines for $n\geq 6$ is very high and
a quantitative analysis is hardly possible. Hence, we concentrate on the $n=3$ and $n=5$
exciton states in the following. 
The theoretical exciton spectra
of these excitons in a magnetic field of $B\leq 3\,\mathrm{T}$ with $\boldsymbol{B}\parallel [001]$
are depicted in Figs.~\ref{fig:Line-spectrum} and~\ref{fig:Line-spectrum2} along with the
exciton states read from experimental data.
From Figs.~\ref{fig:Line-spectrum}(b) and~\ref{fig:Line-spectrum2}(b) 
one can clearly distinguish between
the contribution of the $P$ excitons 
and the $F$ excitons at $B=0\,\mathrm{T}$)~\cite{100}.
It can be seen that the relative oscillator
strength of $F$ excitons significantly increases 
due to state mixing with growing field strength.

By comparing the experimentally observed line splitting to our results,
we can estimate the value of the fourth Luttinger parameter to
\begin{equation}
\kappa=-0.50\pm0.10
\end{equation}
provided that the values listed in Table~\ref{tab:1} are correct 
(see also the according discussion in Ref.~\cite{100}).
Using this value in our numerical calculations, we obtain an
excellent agreement between theory and experiment
for both $n=3$ and $n=5$ excitons in Figs.~\ref{fig:Line-spectrum} 
and~\ref{fig:Line-spectrum2}.
The value of $\kappa$ is further confirmed by the fact that only one
exciton state can be observed at $B=3~\mathrm{T}$
and $E\approx 2.1687~\mathrm{eV}$ in the experiment
for $\sigma^+$ polarized light (see Fig.~\ref{fig:Line-spectrum}).
Only if $-0.52<\kappa<-0.46$
holds, theory predicts two nearly degenerate states. 
For other values of $\kappa$, i.e., for $\kappa>-0.4$ or $\kappa<-0.6$, 
this degeneracy is lifted and two states should 
be observable in the experiment.

We can now use the value of $\kappa=-0.50$ to calculate the
exciton spectra for $\boldsymbol{B}\parallel [110]$ and $\boldsymbol{B}\parallel [111]$.
The results are shown in Figs.~\ref{fig:Line-spectrum3} and~\ref{fig:Line-spectrum4}.
We observe not only an excellent agreement with the experimental results
but also see a clear difference between the spectra for the different orientations
of the magnetic field. This difference is caused only by the cubic part of the
exciton Hamiltonian~[see Eqs.~(\ref{eq:Hh}) and~(\ref{eq:H})].
We also note that the number of exciton states, which can be observed
with $\sigma^+$ and $\sigma^-$ polarized light, differs.
Especially for $\boldsymbol{B}\parallel [110]$ 
$\sigma^{+}$ and $\sigma^{-}$ polarized light belong to the 
same irreducible representation of $C_{\mathrm{2h}}$. Hence,
it is possible to excite a certain exciton state
by $\sigma^+$ \emph{and} by $\sigma^-$ polarized light, for which reason all 20 dipole-allowed 
exciton states can be observed in Fig.~\ref{fig:Line-spectrum3}(a) and in Fig.~\ref{fig:Line-spectrum3}(b).

To compare the theoretically calculated relative oscillator strengths
with the experimental values, we analyze the
experimental spectra using the method of harmonic inversion,
which is presented in detail in Refs.~\cite{HI1,HI2}.
Within the harmonic inversion the spectra
are Fourier transformed to
find the positions $\mathrm{Re}\left(E_k\right)$, widths $\mathrm{Im}\left(E_k\right)$ 
and complex amplitudes $d_k$ of underlying resonances. The spectrum can then 
be expressed by a sum of Lorentzians
\begin{equation}
G\left(E\right)= \mathrm{Im}\left(\sum_{k}\frac{d_k}{E-E_k}\right).\label{eq:Gw}
\end{equation}

The results are presented in Fig.~\ref{fig:hi}.
For $\sigma^+$ and $\sigma^-$ polarized light one can identify
six resonances with exciton states in the theoretical spectrum.
For almost all of these resonances we obtain a very good agreement 
between the modulus $\left|d\right|$ of their amplitudes
and the theoretically calculated relative oscillator strengths.

The harmonic inversion supplies
the true position of the resonances, which is
generally not identical to the position of the transmission minima
due to the asymmetry of the exciton absorption peaks~\cite{GRE,75}. Hence,
we can compare the results for $\mathrm{Re}\left(E_k\right)$
directly to the positions of the exciton states in the theoretical spectrum.
This allows us not only to confirm the value of $\kappa=-0.50\pm 0.10$
but also to determine the first Luttinger parameter more accurately.
The best agreements are obtained for $\gamma_1=1.73\pm0.02$.

\section{Summary and outlook\label{sec:Summary}}

We presented the theory to calculate exciton spectra in Faraday configuration
in a uniform external magnetic field.
Only by taking into account the complex valence band structure
of $\mathrm{Cu_{2}O}$, we could 
obtain an excellent agreement between theory and experiment
as regards not only the relative position but also 
the relative oscillator strengths of the exciton states.
In particular, we showed the significant differences between
the spectra for different orientations of the external magnetic field.
Comparing the theoretical
spectrum for $n=3$ and $n=5$ excitons with experimental results 
and using the method of harmonic inversion, we were able
to determine the fourth Luttinger parameter of cuprous oxide to 
$\kappa=-0.50\pm0.10$.
As a next step, we plan to investigate the spectra of
excitons in $\mathrm{Cu_{2}O}$ in crossed electric and magnetic fields.

\acknowledgments
The Dortmund team acknowledges the support by 
the DFG and the RFBR in the frame of TRR 160.
M.B. acknowledges support by the Russian Ministry of 
Education and Science (Grant No. 14.Z50.31.0021).
We thank also M.~M.~Glazov, 
M.~A.~Semina, H.~Cartarius, J.~Fuchs, and M. Feldmaier
for helpful discussions.

\appendix

\section{Hamiltonian \label{sub:Hamiltonianrm}}

In this section we give the Hamiltonian~(\ref{eq:H}) in terms of irreducible tensors
for the case that the magnetic field is oriented along the $[001]$, $[110]$ or $[111]$ direction~\cite{ED,7_11,44,100}.
Note that we rotate the coordinate system to make the quantization axis or $z$ axis
coincide with the direction of the magnetic field.
Hence, we rotate the coordinate system by the Euler angles
$\left(\alpha,\,\beta,\,\gamma\right)=\left(\pi,\,\pi/2,\,\pi/4\right)$ for
$\boldsymbol{B}\parallel [110]$ and by
$\left(\alpha,\,\beta,\,\gamma\right)=\left(0,\,\arccos(1/\sqrt{3}),\,\pi/4\right)$ for
$\boldsymbol{B}\parallel [111]$.

The first-order and second-order tensor operators
used in the following
correspond, as in Ref.~\cite{44}, to the vector operators 
$\boldsymbol{r}$,~$\boldsymbol{L}=\boldsymbol{r}\times\boldsymbol{p}$,~$\boldsymbol{I}$,~$\boldsymbol{S}_{\mathrm{e/h}}$
and to the second-rank Cartesian operators
\begin{subequations}
\begin{eqnarray}
I_{mn} & = & 3\left\{ I_{m},\, I_{n}\right\} -\delta_{mn}I^{2},\\
P_{mn} & = & 3\left\{ p_{m},\, p_{n}\right\} -\delta_{mn}p^{2},\\
-iM_{mn} & = & 3\left\{ r_{m},\, p_{n}\right\} -\delta_{mn}\boldsymbol{r}\boldsymbol{p},\\
X_{mn} & = & 3\left\{ r_{m},\, r_{n}\right\} -\delta_{mn}r^{2},
\end{eqnarray}
\end{subequations}
respectively. We also use the abbreviation
\begin{equation}
D_{k}^{(2)}=\left[I^{(1)}\times S_{\mathrm{h}}^{(1)}\right]_{k}^{(2)}.
\end{equation}
The coefficients $\gamma'_{1}$, $\mu'$ and $\delta'$
are given by~\cite{7_11,7}
\begin{equation}
\gamma'_{1}=\gamma_{1}+\frac{m_{0}}{m_{\mathrm{e}}},\quad\mu'=\frac{6\gamma_{3}+4\gamma_{2}}{5\gamma'_{1}},\quad\delta'=\frac{\gamma_{3}-\gamma_{2}}{\gamma'_{1}},
\end{equation}
and we define by analogy~\cite{100}
\begin{equation}
\nu=\frac{6\eta_{3}+4\eta_{2}}{5\eta_{1}},\quad\tau=\frac{\eta_{3}-\eta_{2}}{\eta_{1}}.
\end{equation}

Furthermore, we write the Hamiltonian in the form
\begin{equation}
H=H_{0}+\left(eB\right)H_{1}+\left(eB\right)^{2}H_{2}.\label{eq:Hges}
\end{equation}
The expressions for $H_0$, $H_1$ and $H_2$ are given in the following.

\begin{widetext}

\subsection{Magnetic field in [001] direction \label{sub:A001}}

\begin{subequations}
\begin{align}
H_{0} = &\:  E_{\mathrm{g}}-\frac{e^{2}}{4\pi\varepsilon_{0}\varepsilon}\frac{1}{r}+\frac{2}{3}\Delta\left(1+\frac{1}{\hbar^{2}}I^{(1)}\cdot S_{\mathrm{h}}^{(1)}\right)\displaybreak[2]\nonumber \\
\nonumber \\
+ &\: \frac{\gamma_{1}'}{2\hbar^{2}m_{0}}\left\{ \hbar^{2}p^{2}-\frac{\mu'}{3}\left(P^{(2)}\cdot I^{(2)}\right)+\frac{\delta'}{3}\left(\sum_{k=\pm4}\left[P^{(2)}\times I^{(2)}\right]_{k}^{(4)}+\frac{\sqrt{70}}{5}\left[P^{(2)}\times I^{(2)}\right]_{0}^{(4)}\right)\right\}\displaybreak[3]\nonumber \\
\nonumber \\
+ &\: \frac{3\eta_{1}}{\hbar^{2}m_{0}}\left\{ \frac{1}{3}p^{2}\left(I^{(1)}\cdot S_{\mathrm{h}}^{(1)}\right)-\frac{\nu}{3}\left(P^{(2)}\cdot D^{(2)}\right)+\frac{\tau}{3}\left(\sum_{k=\pm4}\left[P^{(2)}\times D^{(2)}\right]_{k}^{(4)}+\frac{\sqrt{70}}{5}\left[P^{(2)}\times D^{(2)}\right]_{0}^{(4)}\right)\right\}
\\
\displaybreak[1]
\nonumber \\
H_{1} = &\:  \frac{1}{m_{\mathrm{e}}}L_{0}^{(1)}+\frac{\mu_{\mathrm{B}}}{e\hbar}\left(g_{c}S_{\mathrm{e}\,0}^{(1)}-g_{s}S_{\mathrm{h}\,0}^{(1)}+\left(3\kappa+\frac{1}{2}g_{s}\right)I_{0}^{(1)}\right)\displaybreak[3]\nonumber \\
\displaybreak[2]\nonumber \\
+  &\: \frac{\gamma_{1}'}{2\hbar^{2}m_{0}}\left\{ -\hbar^{2}L_{0}^{(1)}+\frac{\delta'}{3}\left(\left[M^{(2)}\times I^{(2)}\right]_{-4}^{(4)}-\left[M^{(2)}\times I^{(2)}\right]_{4}^{(4)}\right)\right.\displaybreak[3]\nonumber \\
\nonumber \\
 &\: \qquad\qquad\qquad + \sqrt{\frac{2}{5}}\delta'\left(\left[L^{(1)}\times I^{(2)}\right]_{0}^{(3)}-\frac{1}{3}\left[M^{(2)}\times I^{(2)}\right]_{0}^{(3)}\right)\displaybreak[3]\nonumber \\
\nonumber \\
 &\: \qquad\qquad\qquad + \left.\sqrt{\frac{5}{12}}\mu'\left(\left[L^{(1)}\times I^{(2)}\right]_{0}^{(1)}+\sqrt{\frac{2}{3}}\left[M^{(2)}\times I^{(2)}\right]_{0}^{(1)}\right)\right\} \nonumber \\
\displaybreak[2]\nonumber \\
 + &\: \frac{3\eta_{1}}{\hbar^{2}m_{0}}\left\{ -\frac{1}{3}L_{0}^{(1)}\left(I^{(1)}\cdot S_{\mathrm{h}}^{(1)}\right)+\frac{\tau}{3}\left(\left[M^{(2)}\times D^{(2)}\right]_{-4}^{(4)}-\left[M^{(2)}\times D^{(2)}\right]_{4}^{(4)}\right)\right.\displaybreak[3]\nonumber \\
\nonumber \\
 &\: \qquad\qquad\qquad + \sqrt{\frac{2}{5}}\tau\left(\left[L^{(1)}\times D^{(2)}\right]_{0}^{(3)}-\frac{1}{3}\left[M^{(2)}\times D^{(2)}\right]_{0}^{(3)}\right)\displaybreak[3]\nonumber \\
\nonumber \\
 &\: \qquad\qquad\qquad + \left.\sqrt{\frac{5}{12}}\nu\left(\left[L^{(1)}\times D^{(2)}\right]_{0}^{(1)}+\sqrt{\frac{2}{3}}\left[M^{(2)}\times D^{(2)}\right]_{0}^{(1)}\right)\right\} 
\\
\displaybreak[1]
\nonumber \\
H_{2} = &\: \frac{\gamma_{1}'}{24\hbar^{2}m_{0}}\left\{ \hbar^{2}\left(2r^{2}-\sqrt{\frac{2}{3}}X_{0}^{(2)}\right)-\delta'\left(\sum_{k=\pm4}\left[X^{(2)}\times I^{(2)}\right]_{k}^{(4)}-\frac{2}{\sqrt{70}}\left[X^{(2)}\times I^{(2)}\right]_{0}^{(4)}\right)\right.\displaybreak[3]\nonumber \\
\nonumber \\
 &\: \qquad\qquad\qquad + \left.\frac{\sqrt{14}}{3}\left(\mu'-\frac{12}{35}\delta'\right)\left[X^{(2)}\times I^{(2)}\right]_{0}^{(2)}+\frac{\mu'}{3}\left(X^{(2)}\cdot I^{(2)}\right)+\sqrt{\frac{2}{3}}\left(\mu'-\frac{6}{5}\delta'\right)r^{2}I_{0}^{(2)}\right\} \nonumber \\
\displaybreak[2]\nonumber \\
+ &\: \frac{\eta_{1}}{4\hbar^{2}m_{0}}\left\{ \frac{1}{3}\left(2r^{2}-\sqrt{\frac{2}{3}}X_{0}^{(2)}\right)\left(I^{(1)}\cdot S_{\mathrm{h}}^{(1)}\right)-\tau\left(\sum_{k=\pm4}\left[X^{(2)}\times D^{(2)}\right]_{k}^{(4)}-\frac{2}{\sqrt{70}}\left[X^{(2)}\times D^{(2)}\right]_{0}^{(4)}\right)\right.\displaybreak[3]\nonumber \\
\nonumber \\
 &\: \qquad\qquad\qquad + \left.\frac{\sqrt{14}}{3}\left(\nu-\frac{12}{35}\tau\right)\left[X^{(2)}\times D^{(2)}\right]_{0}^{(2)}+\frac{\nu}{3}\left(X^{(2)}\cdot D^{(2)}\right)+\sqrt{\frac{2}{3}}\left(\nu-\frac{6}{5}\tau\right)r^{2}D_{0}^{(2)}\right\} 
\end{align}
\end{subequations}

\subsection{Magnetic field in [110] direction \label{sub:A110}}

\begin{subequations}
\begin{align}
H_{0} = &\:  E_{\mathrm{g}}-\frac{e^{2}}{4\pi\varepsilon_{0}\varepsilon}\frac{1}{r}+\frac{2}{3}\Delta\left(1+\frac{1}{\hbar^{2}}I^{(1)}\cdot S_{\mathrm{h}}^{(1)}\right)\displaybreak[2]\nonumber \\
\nonumber \\
 &\: +\frac{\gamma_{1}^{'}}{2\hbar^{2}m_{0}}\left\{ \hbar^{2}p^{2}-\frac{\mu^{'}}{3}\left(P^{(2)}\cdot I^{(2)}\right)+\frac{\delta^{'}}{4}\left(\sum_{k=\pm4}\left[P^{(2)}\times I^{(2)}\right]_{k}^{(4)}\right)\right.\displaybreak[3]\nonumber \\
\nonumber \\
 &\: \qquad\qquad\qquad\left.-\frac{\sqrt{7}}{6}\delta^{'}\left(\sum_{k=\pm2}\left[P^{(2)}\times I^{(2)}\right]_{2}^{(4)}+\sqrt{\frac{1}{10}}\left[P^{(2)}\times I^{(2)}\right]_{0}^{(4)}\right)\right\} \nonumber \\
\displaybreak[2]\nonumber \\
 &\: +\frac{3\eta_{1}}{\hbar^{2}m_{0}}\left\{ \frac{1}{3}p^{2}\left(I^{(1)}\cdot S_{\mathrm{h}}^{(1)}\right)-\frac{\nu}{3}\left(P^{(2)}\cdot D^{(2)}\right)+\frac{\tau}{4}\left(\sum_{k=\pm4}\left[P^{(2)}\times D^{(2)}\right]_{k}^{(4)}\right)\right.\displaybreak[3]\nonumber \\
\nonumber \\
 &\: \qquad\qquad\qquad\left.-\frac{\sqrt{7}}{6}\tau\left(\sum_{k=\pm2}\left[P^{(2)}\times D^{(2)}\right]_{k}^{(4)}+\sqrt{\frac{1}{10}}\left[P^{(2)}\times D^{(2)}\right]_{0}^{(4)}\right)\right\} 
\\
\displaybreak[1]
\nonumber \\
H_{1} = &\:  \frac{1}{m_{\mathrm{e}}}L_{0}^{(1)}+\frac{\mu_{\mathrm{B}}}{e\hbar}\left(g_{c}S_{\mathrm{e}\,0}^{(1)}-g_{s}S_{\mathrm{h}\,0}^{(1)}+\left(3\kappa+\frac{1}{2}g_{s}\right)I_{0}^{(1)}\right)\nonumber \\
\displaybreak[2]\nonumber \\
 &\: +\frac{\gamma_{1}^{'}}{2\hbar^{2}m_{0}}\left\{ -\hbar^{2}L_{0}^{(1)}-\frac{\delta^{'}}{16}\left(\sum_{k=\pm4}k\left[M^{(2)}\times I^{(2)}\right]_{k}^{(4)}\right)+\frac{\sqrt{7}}{24}\delta^{'}\left(\sum_{k=\pm2}k\left[M^{(2)}\times I^{(2)}\right]_{k}^{(4)}\right)\right.\displaybreak[3]\nonumber \\
\nonumber \\
 &\: \qquad\qquad\qquad+\frac{1}{4\sqrt{3}}\delta^{'}\left(\sum_{k=\pm2}\left[M^{(2)}\times I^{(2)}\right]_{k}^{(3)}+\sqrt{\frac{2}{15}}\left[M^{(2)}\times I^{(2)}\right]_{0}^{(3)}\right)\displaybreak[3]\nonumber \\
\nonumber \\
 &\: \qquad\qquad\qquad-\frac{\sqrt{3}}{4}\delta^{'}\left(\sum_{k=\pm2}\left[L^{(1)}\times I^{(2)}\right]_{k}^{(3)}+\sqrt{\frac{2}{15}}\left[L^{(1)}\times I^{(2)}\right]_{0}^{(3)}\right)\displaybreak[3]\nonumber \\
\nonumber \\
 &\: \qquad\qquad\qquad\left.+\frac{1}{2}\sqrt{\frac{5}{3}}\mu^{'}\left(\left[L^{(1)}\times I^{(2)}\right]_{0}^{(1)}+\sqrt{\frac{2}{3}}\left[M^{(2)}\times I^{(2)}\right]_{0}^{(1)}\right)\right\} \nonumber \\
\displaybreak[2]\nonumber \\
 &\: +\frac{3\eta_{1}}{\hbar^{2}m_{0}}\left\{ -\frac{1}{3}L_{0}^{(1)}\left(I^{(1)}\cdot S_{\mathrm{h}}^{(1)}\right)-\frac{\tau}{16}\left(\sum_{k=\pm4}k\left[M^{(2)}\times D^{(2)}\right]_{k}^{(4)}\right)+\frac{\sqrt{7}}{24}\tau\left(\sum_{k=\pm2}k\left[M^{(2)}\times D^{(2)}\right]_{k}^{(4)}\right)\right.\displaybreak[3]\nonumber \\
\nonumber \\
 &\: \qquad\qquad\qquad+\frac{1}{4\sqrt{3}}\tau\left(\sum_{k=\pm2}\left[M^{(2)}\times D^{(2)}\right]_{k}^{(3)}+\sqrt{\frac{2}{15}}\left[M^{(2)}\times D^{(2)}\right]_{0}^{(3)}\right)\displaybreak[3]\nonumber \\
\nonumber \\
 &\: \qquad\qquad\qquad-\frac{\sqrt{3}}{4}\tau\left(\sum_{k=\pm2}\left[L^{(1)}\times D^{(2)}\right]_{k}^{(3)}+\sqrt{\frac{2}{15}}\left[L^{(1)}\times D^{(2)}\right]_{0}^{(3)}\right)\displaybreak[3]\nonumber \\
\nonumber \\
 &\: \qquad\qquad\qquad\left.+\frac{1}{2}\sqrt{\frac{5}{3}}\nu\left(\left[L^{(1)}\times D^{(2)}\right]_{0}^{(1)}+\sqrt{\frac{2}{3}}\left[M^{(2)}\times D^{(2)}\right]_{0}^{(1)}\right)\right\} 
\\
\displaybreak[1]
\nonumber \\
H_{2} = &\:  \frac{\gamma_{1}^{'}}{24\hbar^{2}m_{0}}\left\{ \hbar^{2}\left(2r^{2}-\sqrt{\frac{2}{3}}X_{0}^{(2)}\right)-\frac{3}{4}\delta^{'}\left(\sum_{k=\pm4}\left[X^{(2)}\times I^{(2)}\right]_{k}^{(4)}+\frac{1}{3}\sqrt{\frac{2}{35}}\left[X^{(2)}\times I^{(2)}\right]_{0}^{(4)}\right)\right.\displaybreak[3]\nonumber \\
\nonumber \\
 &\: \qquad\qquad\qquad+\frac{\delta^{'}}{2\sqrt{7}}\left(\sum_{k=\pm2}\left[X^{(2)}\times I^{(2)}\right]_{k}^{(4)}\right)+\frac{\delta^{'}}{2\sqrt{3}}\left(\sum_{k=\pm2}k\left[X^{(2)}\times I^{(2)}\right]_{k}^{(3)}\right)\displaybreak[3]\nonumber \\
\nonumber \\
 &\: \qquad\qquad\qquad+\frac{\delta^{'}}{\sqrt{21}}\left(\sum_{k=\pm2}\left[X^{(2)}\times I^{(2)}\right]_{k}^{(2)}\right)+\frac{2}{3}\sqrt{\frac{7}{2}}\left(\mu^{'}+\frac{3}{35}\delta^{'}\right)\left[X^{(2)}\times I^{(2)}\right]_{0}^{(2)}\displaybreak[3]\nonumber \\
\nonumber \\
 &\: \qquad\qquad\qquad\left.+\frac{\mu^{'}}{3}\left(X^{(2)}\cdot I^{(2)}\right)+\sqrt{\frac{2}{3}}\left(\mu^{'}+\frac{3}{10}\delta^{'}\right)r^{2}I_{0}^{(2)}+\frac{\delta^{'}}{2}r^{2}\left(I_{2}^{(2)}+I_{-2}^{(2)}\right)\right\} \nonumber \\
\displaybreak[2]\nonumber \\
 &\: +\frac{\eta_{1}}{4\hbar^{2}m_{0}}\left\{ \frac{1}{3}\left(2r^{2}-\sqrt{\frac{2}{3}}X_{0}^{(2)}\right)\left(I^{(1)}\cdot S_{\mathrm{h}}^{(1)}\right)-\frac{3}{4}\tau\left(\sum_{k=\pm4}\left[X^{(2)}\times D^{(2)}\right]_{k}^{(4)}+\frac{1}{3}\sqrt{\frac{2}{35}}\left[X^{(2)}\times D^{(2)}\right]_{0}^{(4)}\right)\right.\displaybreak[3]\nonumber \\
\nonumber \\
 &\: \qquad\qquad\qquad+\frac{\tau}{2\sqrt{7}}\left(\sum_{k=\pm2}\left[X^{(2)}\times D^{(2)}\right]_{k}^{(4)}\right)+\frac{\tau}{2\sqrt{3}}\left(\sum_{k=\pm2}k\left[X^{(2)}\times D^{(2)}\right]_{k}^{(3)}\right)\displaybreak[3]\nonumber \\
\nonumber \\
 &\: \qquad\qquad\qquad+\frac{\tau}{\sqrt{21}}\left(\sum_{k=\pm2}\left[X^{(2)}\times D^{(2)}\right]_{k}^{(2)}\right)+\frac{2}{3}\sqrt{\frac{7}{2}}\left(\nu+\frac{3}{35}\tau\right)\left[X^{(2)}\times D^{(2)}\right]_{0}^{(2)}\displaybreak[3]\nonumber \\
\nonumber \\
 &\: \qquad\qquad\qquad\left.+\frac{1}{3}\nu\left(X^{(2)}\cdot D^{(2)}\right)+\sqrt{\frac{2}{3}}\left(\nu+\frac{3}{10}\tau\right)r^{2}D_{0}^{(2)}+\frac{\tau}{2}r^{2}\left(D_{2}^{(2)}+D_{-2}^{(2)}\right)\right\} 
\end{align}
\end{subequations}

\subsection{Magnetic field in [111] direction \label{sub:A111}}

\begin{subequations}
\begin{align}
H_{0} = &\:  E_{\mathrm{g}}-\frac{e^{2}}{4\pi\varepsilon_{0}\varepsilon}\frac{1}{r}+\frac{2}{3}\Delta\left(1+\frac{1}{\hbar^{2}}I^{(1)}\cdot S_{\mathrm{h}}^{(1)}\right)\displaybreak[2]\nonumber \\
\nonumber \\
 &\: +\frac{\gamma_{1}^{'}}{2\hbar^{2}m_{0}}\left\{ \hbar^{2}p^{2}-\frac{\mu^{'}}{3}\left(P^{(2)}\cdot I^{(2)}\right)\right.\displaybreak[3]\nonumber \\
\nonumber \\
 &\: \qquad\qquad\qquad\left.+\frac{4}{27}\delta^{'}\left(\sum_{k=\pm3}k\left[P^{(2)}\times I^{(2)}\right]_{k}^{(4)}-3\sqrt{\frac{7}{10}}\left[P^{(2)}\times I^{(2)}\right]_{0}^{(4)}\right)\right\} \nonumber \\
\displaybreak[2]\nonumber \\
 &\: +\frac{3\eta_{1}}{\hbar^{2}m_{0}}\left\{ \frac{1}{3}p^{2}\left(I^{(1)}\cdot S_{\mathrm{h}}^{(1)}\right)-\frac{\nu}{3}\left(P^{(2)}\cdot D^{(2)}\right)\right.\displaybreak[3]\nonumber \\
\nonumber \\
 &\: \qquad\qquad\qquad\left.+\frac{4}{27}\tau\left(\sum_{k=\pm3}k\left[P^{(2)}\times D^{(2)}\right]_{k}^{(4)}-3\sqrt{\frac{7}{10}}\left[P^{(2)}\times D^{(2)}\right]_{0}^{(4)}\right)\right\} 
\\
\displaybreak[1]
\nonumber \\
H_{1} = &\:  \frac{1}{m_{\mathrm{e}}}L_{0}^{(1)}+\frac{\mu_{\mathrm{B}}}{e\hbar}\left(g_{c}S_{\mathrm{e}\,0}^{(1)}-g_{s}S_{\mathrm{h}\,0}^{(1)}+\left(3\kappa+\frac{1}{2}g_{s}\right)I_{0}^{(1)}\right)\nonumber \\
\displaybreak[2]\nonumber \\
 &\: +\frac{\gamma_{1}^{'}}{2\hbar^{2}m_{0}}\left\{ -\hbar^{2}L_{0}^{(1)}-\frac{\delta^{'}}{3}\left(\sum_{k=\pm3}\left[M^{(2)}\times I^{(2)}\right]_{k}^{(4)}\right)\right.\displaybreak[3]\nonumber \\
\nonumber \\
 &\: \qquad\qquad\qquad-\frac{\delta^{'}}{27}\left(\sum_{k=\pm3}k\left[M^{(2)}\times I^{(2)}\right]_{k}^{(3)}-6\sqrt{\frac{2}{5}}\left[M^{(2)}\times I^{(2)}\right]_{0}^{(3)}\right)\displaybreak[3]\nonumber \\
\nonumber \\
 &\: \qquad\qquad\qquad+\frac{\delta^{'}}{9}\left(\sum_{k=\pm3}k\left[L^{(1)}\times I^{(2)}\right]_{k}^{(3)}-6\sqrt{\frac{2}{5}}\left[L^{(1)}\times I^{(2)}\right]_{0}^{(3)}\right)\displaybreak[3]\nonumber \\
\nonumber \\
 &\: \qquad\qquad\qquad\left.+\frac{1}{2}\sqrt{\frac{5}{3}}\mu^{'}\left(\left[L^{(1)}\times I^{(2)}\right]_{0}^{(1)}+\sqrt{\frac{2}{3}}\left[M^{(2)}\times I^{(2)}\right]_{0}^{(1)}\right)\right\} \nonumber \\
\displaybreak[2]\nonumber \\
 &\: +\frac{3\eta_{1}}{\hbar^{2}m_{0}}\left\{ -\frac{1}{3}L_{0}^{(1)}\left(I^{(1)}\cdot S_{\mathrm{h}}^{(1)}\right)-\frac{\tau}{3}\left(\sum_{k=\pm3}\left[M^{(2)}\times D^{(2)}\right]_{k}^{(4)}\right)\right.\displaybreak[3]\nonumber \\
\nonumber \\
 &\: \qquad\qquad\qquad-\frac{\tau}{27}\left(\sum_{k=\pm3}k\left[M^{(2)}\times D^{(2)}\right]_{k}^{(3)}-6\sqrt{\frac{2}{5}}\left[M^{(2)}\times D^{(2)}\right]_{0}^{(3)}\right)\displaybreak[3]\nonumber \\
\nonumber \\
 &\: \qquad\qquad\qquad+\frac{\tau}{9}\left(\sum_{k=\pm3}k\left[L^{(1)}\times D^{(2)}\right]_{k}^{(3)}-6\sqrt{\frac{2}{5}}\left[L^{(1)}\times D^{(2)}\right]_{0}^{(3)}\right)\displaybreak[3]\nonumber \\
\nonumber \\
 &\: \qquad\qquad\qquad\left.+\frac{1}{2}\sqrt{\frac{5}{3}}\nu\left(\left[L^{(1)}\times D^{(2)}\right]_{0}^{(1)}+\sqrt{\frac{2}{3}}\left[M^{(2)}\times D^{(2)}\right]_{0}^{(1)}\right)\right\} 
\\
\displaybreak[1]
\nonumber \\
H_{2} = &\:  \frac{\gamma_{1}^{'}}{24\hbar^{2}m_{0}}\left\{ \hbar^{2}\left(2r^{2}-\sqrt{\frac{2}{3}}X_{0}^{(2)}\right)+\frac{2}{3}\delta^{'}\left(-\frac{1}{3}\sum_{k=\pm3}k\left[X^{(2)}\times I^{(2)}\right]_{k}^{(4)}-\sqrt{\frac{2}{35}}\left[X^{(2)}\times I^{(2)}\right]_{0}^{(4)}\right)\right.\displaybreak[3]\nonumber \\
\nonumber \\
 &\: \qquad\qquad\qquad-\frac{2}{3}\delta^{'}\sum_{k=\pm3}\left[X^{(2)}\times I^{(2)}\right]_{k}^{(3)}+\frac{2}{3}\sqrt{\frac{7}{2}}\left(\mu^{'}+\frac{8}{35}\delta^{'}\right)\left[X^{(2)}\times I^{(2)}\right]_{0}^{(2)}\displaybreak[3]\nonumber \\
\nonumber \\
 &\: \qquad\qquad\qquad\left.+\frac{\mu^{'}}{3}\left(X^{(2)}\cdot I^{(2)}\right)+\sqrt{\frac{2}{3}}\left(\mu^{'}+\frac{4}{5}\delta^{'}\right)r^{2}I_{0}^{(2)}\right\} \displaybreak[2]\nonumber \\
 &\: +\frac{\eta_{1}}{4\hbar^{2}m_{0}}\left\{ \frac{1}{3}\left(2r^{2}-\sqrt{\frac{2}{3}}X_{0}^{(2)}\right)\left(I^{(1)}\cdot S_{\mathrm{h}}^{(1)}\right)+\frac{2}{3}\tau\left(-\frac{1}{3}\sum_{k=\pm3}k\left[X^{(2)}\times I^{(2)}\right]_{k}^{(4)}-\sqrt{\frac{2}{35}}\left[X^{(2)}\times D^{(2)}\right]_{0}^{(4)}\right)\right.\displaybreak[3]\nonumber \\
\nonumber \\
 &\: \qquad\qquad\qquad-\frac{2}{3}\tau\sum_{k=\pm3}\left[X^{(2)}\times I^{(2)}\right]_{k}^{(3)}+\frac{2}{3}\sqrt{\frac{7}{2}}\left(\nu+\frac{8}{35}\tau\right)\left[X^{(2)}\times D^{(2)}\right]_{0}^{(2)}\displaybreak[3]\nonumber \\
\nonumber \\
 &\: \qquad\qquad\qquad\left.+\frac{\nu}{3}\left(X^{(2)}\cdot D^{(2)}\right)+\sqrt{\frac{2}{3}}\left(\nu+\frac{4}{5}\tau\right)r^{2}D_{0}^{(2)}\right\} 
\end{align}
\end{subequations}

\section{Oscillator strengths \label{sub:Oscillator-strengths}}

We now give the formula for the expression
\begin{equation}
\lim_{r\rightarrow0}\frac{\partial}{\partial r}\,_{D}\left\langle 2,\,M'_{F_t}\middle|\Psi\left(\boldsymbol{r}\right)\right\rangle,
\end{equation}
which is needed for the evaluation of the relative 
oscillator strength $f_{\mathrm{rel}}$~(\ref{eq:frel}).
Using the wave function of Eq.~(\ref{eq:ansatz}), we find
\begin{align}
\lim_{r\rightarrow0}\frac{\partial}{\partial r} \,_{D}\left\langle 2,\,M'_{F_t}\middle|\Psi\left(\boldsymbol{r}\right)\right\rangle= &\: \sum_{NJFF_{t}}\sum_{M_{S_{\mathrm{e}}}M_{I}}c_{N1JFF_{t}M'_{F_t}}\;\frac{1}{3}\sqrt{\frac{10}{\alpha^5}}\left(-1\right)^{F-J-3M_{S_{\mathrm{e}}}-M_{I}+3M'_{F_t}+\frac{3}{2}}\displaybreak[3]\nonumber \\
\nonumber \\
\times  &\:  \left[(N+1)(N+3)(2J+1)(2F+1)(2F_{t}+1)\right]^{\frac{1}{2}}\displaybreak[3]\nonumber \\
\nonumber \\
\times  &\:  \left(\begin{array}{ccc}
F & \frac{1}{2} & F_{t}\\
M'_{F_t}-M_{S_{\mathrm{e}}} & M_{S_{\mathrm{e}}} & -M'_{F_t}
\end{array}\right)\left(\begin{array}{ccc}
1 & J & F\\
M'_{F_t}-M_{I} & M_{I}-M_{S_{\mathrm{e}}} & M_{S_{\mathrm{e}}}-M'_{F_t}
\end{array}\right)\displaybreak[3]\nonumber \\
\nonumber \\
\times  &\:  \left(\begin{array}{ccc}
1 & \frac{1}{2} & J\\
M_{I} & -M_{S_{\mathrm{e}}} & M_{S_{\mathrm{e}}}-M_{I}
\end{array}\right)\left(\begin{array}{ccc}
1 & 1 & 2\\
M_{I} & M'_{F_t}-M_{I} & -M'_{F_t}
\end{array}\right).\label{eq:osc_str}
\end{align}

\section{Matrix elements \label{sub:Matrix-elements}}

In this section we give the matrix elements of the terms of the Hamiltonian
$H$ [Eq.~(\ref{eq:Hges})] in the basis of Eq.~(\ref{eq:ansatz})
in Hartree units using the formalism of irreducible tensors~\cite{ED}.
The matrix elements of the Hamiltonian $H_{0}$ [Eq.~(\ref{eq:Hges})]
are given in the Appendix of Ref.~\cite{100}. We use the abbreviation
\begin{equation}
\tilde{\delta}_{\Pi\Pi'}=\delta_{LL'}\delta_{JJ'}\delta_{FF'}\delta_{F_{t}F'_{t}}\delta_{M_{F_{t}}M_{F_{t}}'}
\end{equation}
in the following. The functions of the form $\left(R_{1}\right)_{nL}^{j}$
are taken from the recursion relations of the Coulomb-Sturmian functions
in the Appendix of Ref.~\cite{100}.

\begin{align}
\left\langle \Pi'\left|r^{2}\right|\Pi\right\rangle = &\: \tilde{\delta}_{\Pi\Pi'}\sum_{j=-3}^{3}\left(R_{3}\right)_{NL}^{j}\left[N+L+j+1\right]^{-1}\delta_{N',N+j}
\\
\displaybreak[1]
\nonumber \\
\left\langle \Pi'\left|r^{2}I_{q}^{(2)}\right|\Pi\right\rangle = &\:  \delta_{LL'}3\sqrt{5}\left(-1\right)^{F'_{t}+F_{t}-M_{F_{t}}'+2F'+L+2J}\displaybreak[3]\nonumber \\
\nonumber \\
\times &\: \left[\left(2F_{t}+1\right)\left(2F'_{t}+1\right)\left(2F+1\right)\left(2F'+1\right)\left(2J+1\right)\left(2J'+1\right)\right]^{\frac{1}{2}}\displaybreak[3]\nonumber \\
\nonumber \\
\times &\: \left(\begin{array}{ccc}
F'_{t} & 2 & F_{t}\\
-M_{F_{t}}' & q & M_{F_{t}}
\end{array}\right)\left\{ \begin{array}{ccc}
F' & F'_{t} & \frac{1}{2}\\
F_{t} & F & 2
\end{array}\right\} \left\{ \begin{array}{ccc}
J' & F' & L\\
F & J & 2
\end{array}\right\} \left\{ \begin{array}{ccc}
1 & J' & \frac{1}{2}\\
J & 1 & 2
\end{array}\right\}\displaybreak[3]\nonumber \\
\nonumber \\
\times &\: \sum_{j=-3}^{3}\left(R_{3}\right)_{NL}^{j}\left[N+L+j+1\right]^{-1}\delta_{N',N+j}
\\
\displaybreak[1]
\nonumber \\
\left\langle \Pi'\left|X_{q}^{(2)}\right|\Pi\right\rangle =  &\: \delta_{JJ'}\left(-1\right)^{F'_{t}+F_{t}-M_{F_{t}}'+F'+F+L'+J+\frac{1}{2}}\left\langle N'\, L'\left\Vert X^{(2)}\right\Vert N\, L\right\rangle\displaybreak[3]\nonumber \\
\nonumber \\
\times &\: \left[\left(2F_{t}+1\right)\left(2F'_{t}+1\right)\left(2F+1\right)\left(2F'+1\right)\right]^{\frac{1}{2}}\displaybreak[3]\nonumber \\
\nonumber \\
\times &\: \left(\begin{array}{ccc}
F'_{t} & 2 & F_{t}\\
-M_{F_{t}}' & q & M_{F_{t}}
\end{array}\right)\left\{ \begin{array}{ccc}
F' & F'_{t} & \frac{1}{2}\\
F_{t} & F & 2
\end{array}\right\} \left\{ \begin{array}{ccc}
L' & F' & J\\
F & L & 2
\end{array}\right\} 
\\
\displaybreak[1]
\nonumber \\
\left\langle \Pi'\left|S_{\mathrm{e}\,0}^{(1)}\right|\Pi\right\rangle = &\: \delta_{LL'}\delta_{JJ'}\delta_{FF'}\delta_{M_{F_{t}}M'_{F_{t}}}\sqrt{\frac{3}{2}}\left(-1\right)^{2F'_{t}-M_{F_{t}}+F+\frac{3}{2}}\displaybreak[3]\nonumber \\
\nonumber \\
\times &\: \left[\left(2F_{t}+1\right)\left(2F'_{t}+1\right)\right]^{\frac{1}{2}}\left(\begin{array}{ccc}
F'_{t} & 1 & F_{t}\\
-M_{F_{t}} & 0 & M_{F_{t}}
\end{array}\right)\left\{ \begin{array}{ccc}
\frac{1}{2} & F'_{t} & F\\
F_{t} & \frac{1}{2} & 1
\end{array}\right\}\displaybreak[3]\nonumber \\
\nonumber \\
\times  &\:  \sum_{j=-1}^{1}\left(R_{1}\right)_{NL}^{j}\left[N+L+j+1\right]^{-1}\delta_{N',N+j}
\\
\displaybreak[1]
\nonumber \\
\left\langle \Pi'\left|S_{\mathrm{h}\,0}^{(1)}\right|\Pi\right\rangle = &\: \delta_{LL'}\delta_{M_{F_{t}}M'_{F_{t}}}\sqrt{\frac{3}{2}}\left(-1\right)^{F'_{t}+F_{t}-M_{F_{t}}+2F'+L+J+J'+1}\displaybreak[3]\nonumber \\
\nonumber \\
\times &\: \left[\left(2F_{t}+1\right)\left(2F'_{t}+1\right)\left(2F+1\right)\left(2F'+1\right)\left(2J+1\right)\left(2J'+1\right)\right]^{\frac{1}{2}}\displaybreak[3]\nonumber \\
\nonumber \\
\times &\: \left(\begin{array}{ccc}
F'_{t} & 1 & F_{t}\\
-M_{F_{t}} & 0 & M_{F_{t}}
\end{array}\right)\left\{ \begin{array}{ccc}
F' & F'_{t} & \frac{1}{2}\\
F_{t} & F & 1
\end{array}\right\} \left\{ \begin{array}{ccc}
J' & F' & L\\
F & J & 1
\end{array}\right\} \left\{ \begin{array}{ccc}
\frac{1}{2} & J' & 1\\
J & \frac{1}{2} & 1
\end{array}\right\} \displaybreak[3]\nonumber \\
\nonumber \\
\times  &\: \sum_{j=-1}^{1}\left(R_{1}\right)_{NL}^{j}\left[N+L+j+1\right]^{-1}\delta_{N',N+j}
\\
\displaybreak[1]
\nonumber \\
\left\langle \Pi'\left|L_{0}^{(1)}\right|\Pi\right\rangle = &\:  \delta_{LL'}\delta_{JJ'}\delta_{M_{F_{t}}M'_{F_{t}}}\left(-1\right)^{F'_{t}+F_{t}-M_{F_{t}}+F+F'+L+J+\frac{1}{2}}\displaybreak[3]\nonumber \\
\nonumber \\
\times &\: \left[\left(2F_{t}+1\right)\left(2F'_{t}+1\right)\left(2F+1\right)\left(2F'+1\right)L\left(L+1\right)\left(2L+1\right)\right]^{\frac{1}{2}}\displaybreak[3]\nonumber \\
\nonumber \\
\times  &\: \left(\begin{array}{ccc}
F'_{t} & 1 & F_{t}\\
-M_{F_{t}} & 0 & M_{F_{t}}
\end{array}\right)\left\{ \begin{array}{ccc}
F' & F'_{t} & \frac{1}{2}\\
F_{t} & F & 1
\end{array}\right\} \left\{ \begin{array}{ccc}
L & F' & J\\
F & L & 1
\end{array}\right\}\displaybreak[3]\nonumber \\
\nonumber \\
\times &\: \sum_{j=-1}^{1}\left(R_{1}\right)_{NL}^{j}\left[N+L+j+1\right]^{-1}\delta_{N',N+j}
\\
\displaybreak[1]
\nonumber \\
\left\langle \Pi'\left|I_{0}^{(1)}\right|\Pi\right\rangle = &\: \delta_{LL'}\delta_{M_{F_{t}}M'_{F_{t}}}\sqrt{6}\left(-1\right)^{F'_{t}+F_{t}-M_{F_{t}}+2F'+L+2J+1}\displaybreak[3]\nonumber \\
\nonumber \\
\times &\: \left[\left(2F_{t}+1\right)\left(2F'_{t}+1\right)\left(2F+1\right)\left(2F'+1\right)\left(2J+1\right)\left(2J'+1\right)\right]^{\frac{1}{2}}\displaybreak[3]\nonumber \\
\nonumber \\
\times &\: \left(\begin{array}{ccc}
F'_{t} & 1 & F_{t}\\
-M_{F_{t}} & 0 & M_{F_{t}}
\end{array}\right)\left\{ \begin{array}{ccc}
F' & F'_{t} & \frac{1}{2}\\
F_{t} & F & 1
\end{array}\right\} \left\{ \begin{array}{ccc}
J' & F' & L\\
F & J & 1
\end{array}\right\} \left\{ \begin{array}{ccc}
1 & J' & \frac{1}{2}\\
J & 1 & 1
\end{array}\right\} \displaybreak[3]\nonumber \\
\nonumber \\
\times &\: \sum_{j=-1}^{1}\left(R_{1}\right)_{NL}^{j}\left[N+L+j+1\right]^{-1}\delta_{N',N+j}
\\
\displaybreak[1]
\nonumber \\
\left\langle \Pi'\left|X^{(2)}\cdot I^{(2)}\right|\Pi\right\rangle =  &\: \sqrt{5}\left[X^{(2)}\times I^{(2)}\right]_{0}^{(0)}
\\
\displaybreak[1]
\nonumber \\
\left\langle \Pi'\left|\left[M^{(2)}\times I^{(2)}\right]_{q}^{(K)}\right|\Pi\right\rangle = &\: 3\sqrt{5}\left(-1\right)^{F'_{t}+F_{t}-M'_{F_{t}}+F'+J+K}\left\langle N'\, L'\left\Vert M^{(2)}\right\Vert N\, L\right\rangle\displaybreak[3]\nonumber \\
\nonumber \\
\times &\: \left[\left(2F_{t}+1\right)\left(2F'_{t}+1\right)\left(2F+1\right)\left(2F'+1\right)\left(2K+1\right)\left(2J+1\right)\left(2J'+1\right)\right]^{\frac{1}{2}}\displaybreak[3]\nonumber \\
\nonumber \\
\times &\: \left(\begin{array}{ccc}
F'_{t} & K & F_{t}\\
-M'_{F_{t}} & q & M_{F_{t}}
\end{array}\right)\left\{ \begin{array}{ccc}
F' & F'_{t} & \frac{1}{2}\\
F_{t} & F & K
\end{array}\right\} \left\{ \begin{array}{ccc}
1 & J' & \frac{1}{2}\\
J & 1 & 2
\end{array}\right\}\left\{ \begin{array}{ccc}
L' & L & 2\\
J' & J & 2\\
F' & F & K
\end{array}\right\}  
\\
\displaybreak[1]
\nonumber \\
\left\langle \Pi'\left|\left[L^{(1)}\times I^{(2)}\right]_{q}^{(K)}\right|\Pi\right\rangle =  &\: \delta_{LL'}3\sqrt{5}\left(-1\right)^{F'_{t}+F_{t}-M_{F_{t}}'+F'+J+K}\left[L\left(L+1\right)\left(2L+1\right)\right]^{\frac{1}{2}}\displaybreak[3]\nonumber \\
\nonumber \\
\times &\: \left[\left(2F_{t}+1\right)\left(2F'_{t}+1\right)\left(2F+1\right)\left(2F'+1\right)\left(2K+1\right)\left(2J+1\right)\left(2J'+1\right)\right]^{\frac{1}{2}}\displaybreak[3]\nonumber \\
\nonumber \\
\times &\: \left(\begin{array}{ccc}
F'_{t} & K & F_{t}\\
-M_{F_{t}}' & q & M_{F_{t}}
\end{array}\right)\left\{ \begin{array}{ccc}
F' & F'_{t} & \frac{1}{2}\\
F_{t} & F & K
\end{array}\right\} \left\{ \begin{array}{ccc}
1 & J' & \frac{1}{2}\\
J & 1 & 2
\end{array}\right\}\left\{ \begin{array}{ccc}
L' & L & 1\\
J' & J & 2\\
F' & F & K
\end{array}\right\} \displaybreak[3]\nonumber \\
\nonumber \\
\times &\:  \sum_{j=-1}^{1}\left(R_{1}\right)_{NL}^{j}\left[N+L+j+1\right]^{-1}\delta_{N',N+j}
\\
\displaybreak[1]
\nonumber \\
\left\langle \Pi'\left|\left[X^{(2)}\times I^{(2)}\right]_{q}^{(K)}\right|\Pi\right\rangle = &\: 3\sqrt{5}\left(-1\right)^{F'_{t}+F_{t}-M'_{F_{t}}+F'+J+K}\left\langle N'\, L'\left\Vert X^{(2)}\right\Vert N\, L\right\rangle \displaybreak[3]\nonumber \\
\nonumber \\
\times &\: \left[\left(2F_{t}+1\right)\left(2F'_{t}+1\right)\left(2F+1\right)\left(2F'+1\right)\left(2K+1\right)\left(2J+1\right)\left(2J'+1\right)\right]^{\frac{1}{2}}\displaybreak[3]\nonumber \\
\nonumber \\
\times &\: \left(\begin{array}{ccc}
F'_{t} & K & F_{t}\\
-M'_{F_{t}} & q & M_{F_{t}}
\end{array}\right)\left\{ \begin{array}{ccc}
F' & F'_{t} & \frac{1}{2}\\
F_{t} & F & K
\end{array}\right\} \left\{ \begin{array}{ccc}
1 & J' & \frac{1}{2}\\
J & 1 & 2
\end{array}\right\}\left\{ \begin{array}{ccc}
L' & L & 2\\
J' & J & 2\\
F' & F & K
\end{array}\right\} 
\\
\displaybreak[1]
\nonumber \\
\left\langle \Pi'\left|L_{0}^{(1)}\left(I^{(1)}\cdot S_{\mathrm{h}}^{(1)}\right)\right|\Pi\right\rangle = &\: \frac{1}{2}\left(J\left(J+1\right)-\frac{11}{4}\right)\left\langle \Pi'\left|L_{0}^{(1)}\right|\Pi\right\rangle 
\\
\displaybreak[1]
\nonumber \\
\left\langle \Pi'\left|r^{2}\left(I^{(1)}\cdot S_{\mathrm{h}}^{(1)}\right)\right|\Pi\right\rangle = &\: \frac{1}{2}\left(J\left(J+1\right)-\frac{11}{4}\right)\left\langle \Pi'\left|r^{2}\right|\Pi\right\rangle 
\\
\displaybreak[1]
\nonumber \\
\left\langle \Pi'\left|X_{q}^{(2)}\left(I^{(1)}\cdot S_{\mathrm{h}}^{(1)}\right)\right|\Pi\right\rangle = &\: \frac{1}{2}\left(J\left(J+1\right)-\frac{11}{4}\right)\left\langle \Pi'\left|X_{q}^{(2)}\right|\Pi\right\rangle 
\\
\displaybreak[1]
\nonumber \\
\left\langle \Pi'\left|r^{2}\left[I^{(1)}\times S_{\mathrm{h}}^{(1)}\right]_{q}^{(2)}\right|\Pi\right\rangle = &\: \delta_{LL'}3\sqrt{5}\left(-1\right)^{F_{t}'+F_{t}-M_{F_{t}}'+2F'+L+J+\frac{1}{2}}\displaybreak[3]\nonumber \\
\nonumber \\
\times &\: \left[(2F_{t}+1)(2F_{t}'+1)(2F+1)(2F'+1)(2J+1)(2J'+1)\right]^{\frac{1}{2}}\displaybreak[3]\nonumber \\
\nonumber \\
\times &\: \left(\begin{array}{ccc}
F_{t}' & 2 & F_{t}\\
-M_{F_{t}}' & q & M_{F_{t}}
\end{array}\right)\left\{ \begin{array}{ccc}
F' & F_{t}' & \frac{1}{2}\\
F_{t} & F & 2
\end{array}\right\} \left\{ \begin{array}{ccc}
J' & F' & L\\
F & J & 2
\end{array}\right\} \left\{ \begin{array}{ccc}
1 & 1 & 1\\
\frac{1}{2} & \frac{1}{2} & 1\\
J' & J & 2
\end{array}\right\}\displaybreak[3]\nonumber \\
\nonumber \\
\times &\: \sum_{j=-3}^{3}\left(R_{3}\right)_{NL}^{j}\left[N+L+j+1\right]^{-1}\delta_{N',N+j}
\\
\displaybreak[1]
\nonumber \\
\left\langle \Pi'\left|X^{(2)}\cdot\left[I^{(1)}\times S_{\mathrm{h}}^{(1)}\right]^{(2)}\right|\Pi\right\rangle = &\: \sqrt{5}\left\langle \Pi'\left|\left[X^{(2)}\times\left[I^{(1)}\times S_{\mathrm{h}}^{(1)}\right]^{(2)}\right]_{0}^{(0)}\right|\Pi\right\rangle 
\\
\displaybreak[1]
\nonumber \\
\left\langle \Pi'\left|\left[X^{(2)}\times\left[I^{(1)}\times S_{\mathrm{h}}^{(1)}\right]^{(2)}\right]_{q}^{(K)}\right|\Pi\right\rangle = &\: 3\sqrt{5}\left(-1\right)^{F_{t}'+F_{t}-M_{F_{t}}'+F'+K+\frac{1}{2}}\left\langle N'\, L'\left\Vert X^{(2)}\right\Vert N\, L\right\rangle \displaybreak[3]\nonumber \\
\nonumber \\
\times &\: \left[(2F_{t}+1)(2F_{t}'+1)(2F+1)(2F'+1)(2K+1)(2J+1)(2J'+1)\right]^{\frac{1}{2}}\displaybreak[3]\nonumber \\
\nonumber \\
\times &\: \left(\begin{array}{ccc}
F_{t}' & K & F_{t}\\
-M_{F_{t}}' & q & M_{F_{t}}
\end{array}\right)\left\{ \begin{array}{ccc}
F' & F_{t}' & \frac{1}{2}\\
F_{t} & F & K
\end{array}\right\} \left\{ \begin{array}{ccc}
L' & L & 2\\
J' & J & 2\\
F' & F & K
\end{array}\right\} \left\{ \begin{array}{ccc}
1 & 1 & 1\\
\frac{1}{2} & \frac{1}{2} & 1\\
J' & J & 2
\end{array}\right\} 
\\
\displaybreak[1]
\nonumber \\
\left\langle \Pi'\left|\left[M^{(2)}\times\left[I^{(1)}\times S_{\mathrm{h}}^{(1)}\right]^{(2)}\right]_{q}^{(K)}\right|\Pi\right\rangle = &\: 3\sqrt{5}\left(-1\right)^{F_{t}'+F_{t}-M_{F_{t}}'+F'+K+\frac{1}{2}}\left\langle N'\, L'\left\Vert M^{(2)}\right\Vert N\, L\right\rangle \displaybreak[3]\nonumber \\
\nonumber \\
\times &\: \left[(2F_{t}+1)(2F_{t}'+1)(2F+1)(2F'+1)(2K+1)(2J+1)(2J'+1)\right]^{\frac{1}{2}}\displaybreak[3]\nonumber \\
\nonumber \\
\times &\: \left(\begin{array}{ccc}
F_{t}' & K & F_{t}\\
-M_{F_{t}}' & q & M_{F_{t}}
\end{array}\right)\left\{ \begin{array}{ccc}
F' & F_{t}' & \frac{1}{2}\\
F_{t} & F & K
\end{array}\right\} \left\{ \begin{array}{ccc}
L' & L & 2\\
J' & J & 2\\
F' & F & K
\end{array}\right\} \left\{ \begin{array}{ccc}
1 & 1 & 1\\
\frac{1}{2} & \frac{1}{2} & 1\\
J' & J & 2
\end{array}\right\} 
\\
\displaybreak[1]
\nonumber \\
\left\langle \Pi'\left|\left[L^{(1)}\times\left[I^{(1)}\times S_{\mathrm{h}}^{(1)}\right]^{(2)}\right]_{q}^{(K)}\right|\Pi\right\rangle = &\: \delta_{LL'}3\sqrt{5}\left(-1\right)^{F_{t}'+F_{t}-M_{F_{t}}'+F'+K+\frac{1}{2}}\left[L(L+1)(2L+1)\right]^{\frac{1}{2}}\displaybreak[3]\nonumber \\
\nonumber \\
\times &\: \left[(2F_{t}+1)(2F_{t}'+1)(2F+1)(2F'+1)(2K+1)(2J+1)(2J'+1)\right]^{\frac{1}{2}}\displaybreak[3]\nonumber \\
\nonumber \\
\times &\: \left(\begin{array}{ccc}
F_{t}' & K & F_{t}\\
-M_{F_{t}}' & q & M_{F_{t}}
\end{array}\right)\left\{ \begin{array}{ccc}
F' & F_{t}' & \frac{1}{2}\\
F_{t} & F & K
\end{array}\right\} \left\{ \begin{array}{ccc}
L' & L & 2\\
J' & J & 2\\
F' & F & K
\end{array}\right\} \left\{ \begin{array}{ccc}
1 & 1 & 1\\
\frac{1}{2} & \frac{1}{2} & 1\\
J' & J & 2
\end{array}\right\}\displaybreak[3]\nonumber \\
\nonumber \\
\times &\: \sum_{j=-1}^{1}\left(R_{1}\right)_{NL}^{j}\left[N+L+j+1\right]^{-1}\delta_{N',N+j}
\end{align}

\section{Reduced matrix elements \label{sub:Reduced-matrix-elements}}

We now list the values of the reduced matrix elements of the form
$\left\langle N'\, L'\left\Vert A^{(j)}\right\Vert N\, L\right\rangle $.
The functions of the form $\left(R_{1}\right)_{NL}^{j}$ and the integral
$I_{N'\, L';N\, L}$ are taken from the Appendix of Ref.~\cite{100}.

\begin{align}
\left\langle N'\, L'\left\Vert X^{(2)}\right\Vert N\, L\right\rangle = &\: \delta_{L',L+2}\,\frac{3}{2}\frac{1}{\left(L+1\right)\left(L+2\right)}\left[\prod_{j=1}^{5}\left(2L+j\right)\right]^{\frac{1}{2}}\left[\sum_{j=-5}^{1}\frac{\left(R_{1}LN_{2}\right)_{NL0}^{j\,2}}{(N+L+j+3)}\,\delta_{N',N+j}\right]\nonumber \\
\displaybreak[2]\nonumber \\
+ &\: \delta_{L',L}\,\left(-\sqrt{\frac{3}{8}}\right)\frac{1}{L\left(L+1\right)}\left[\prod_{j=-1}^{3}\left(2L+j\right)\right]^{\frac{1}{2}}\left[\sum_{j=-3}^{3}\frac{3\left(R_{1}LN_{2}\right)_{NL0}^{j\,0}-\left(R_{3}\right)_{NL}^{j}}{(N+L+j+1)}\,\delta_{N',N+j}\right]\nonumber \\
\displaybreak[2]\nonumber \\
+ &\: \delta_{L',L-2}\,\frac{3}{2}\frac{1}{L\left(L-1\right)}\left[\prod_{j=-3}^{1}\left(2L+j\right)\right]^{\frac{1}{2}}\left[\sum_{j=-1}^{5}\frac{\left(R_{1}LN_{2}\right)_{NL0}^{j\,-2}}{(N+L+j-1)}\,\delta_{N',N+j}\right]
\\
\displaybreak[1]
\nonumber \\
\left\langle N'\, L'\left\Vert M^{(2)}\right\Vert N\, L\right\rangle = &\: \delta_{L',L+2}\,\frac{3}{2}\left[\frac{\left(2L+4\right)\left(2L+2\right)}{\left(2L+3\right)}\right]^{\frac{1}{2}}\displaybreak[3]\nonumber \\
\nonumber \\
&\: \times \left[\sum_{j=-3}^{3}\left(R_{3}P_{1}\right)_{NL}^{j}I_{N'\, L+2;\, N+j\, L}+\sum_{j=-2}^{2}\left(-L\left(R_{2}\right)_{NL}^{j}\right)I_{N'\, L+2;\, N+j\, L}\right]\nonumber \\
\displaybreak[2]\nonumber \\
+ &\: \delta_{L',L}\,\left(-\frac{\sqrt{3}}{2}\right)\left[\frac{L\left(2L+1\right)\left(2L+2\right)}{\left(2L+3\right)\left(2L-1\right)}\right]^{\frac{1}{2}}\displaybreak[3]\nonumber \\
\nonumber \\
&\: \times \left[\sum_{j=-2}^{2}\frac{2\left(R_{2}P_{1}\right)_{NL}^{j}}{(N+L+j+1)}\,\delta_{N',N+j}+\sum_{j=-1}^{1}\frac{3\left(R_{1}\right)_{NL}^{j}}{(N+L+j+1)}\,\delta_{N',N+j}\right]\nonumber \\
\displaybreak[2]\nonumber \\
+ &\: \delta_{L',L-2}\,\frac{3}{2}\left[\frac{\left(2L\right)\left(2L-2\right)}{\left(2L-1\right)}\right]^{\frac{1}{2}}\displaybreak[3]\nonumber \\
\nonumber \\
&\: \times \left[\sum_{j=-3}^{3}\left(R_{3}P_{1}\right)_{NL}^{j}I_{N'\, L-2;\, N+j\, L}+\sum_{j=-2}^{2}\left(L+1\right)\left(R_{2}\right)_{NL}^{j}I_{N'\, L-2;\, N+j\, L}\right]
\end{align}


\end{widetext}





\end{document}